\documentclass{aa}

\usepackage[pdfpagelabels=false]{hyperref}
\hypersetup{colorlinks=true,
            linkcolor=blue,
            citecolor=blue,
            filecolor=blue,
            urlcolor=blue,
            }

\usepackage{graphicx}	
\usepackage{amsmath}	
\usepackage{amssymb}	
\usepackage{multicol}        
\usepackage{bm}		
\usepackage{pdflscape}	
\usepackage{xcolor}

\newcommand{\hGpc}{{\ifmmode{h^{-1}{\rm Gpc}}\else{$h^{-1}$Gpc}\fi}}
\newcommand{\hMpc}{{\ifmmode{h^{-1}{\rm Mpc}}\else{$h^{-1}$Mpc}\fi}}
\newcommand{\hkpc}{{\ifmmode{h^{-1}{\rm kpc}}\else{$h^{-1}$kpc}\fi}}
\newcommand{\hMsun}{{\ifmmode{h^{-1}{\rm {M_{\odot}}}}\else{$h^{-1}{\rm{M_{\odot}}}$}\fi}}
\newcommand{\Mstar}{{\ifmmode{M_{*}}\else{$M_{*}$}\fi}}
\newcommand{\Mhalo}{{\ifmmode{M_{\rm Halo}}\else{$M_{\rm Halo}$}\fi}}
\newcommand{\Ngal}{{\ifmmode{N_{\rm gal}}\else{$N_{\rm gal}$}\fi}}
\newcommand{\Norph}{{\ifmmode{N_{\rm orphan}}\else{$N_{\rm orphan}$}\fi}}
\newcommand{\Nxorph}{{\ifmmode{N_{\rm non-orphan}}\else{$N_{\rm non-orphan}$}\fi}}
\newcommand{\Zsolar}{{\ifmmode{Z_{\odot}}\else{$Z_{\odot}$}\fi}}
\newcommand{\Msun}{{\ifmmode{{\rm {M_{\odot}}}}\else{${\rm{M_{\odot}}}$}\fi}}
\newcommand{\ltsima}{$\; \buildrel < \over \sim \;$}
\newcommand{\gtsima}{$\; \buildrel > \over \sim \;$}
\newcommand{\lsim}{\lower.5ex\hbox{\ltsima}}
\newcommand{\gsim}{\lower.5ex\hbox{\gtsima}}

\newcommand{\Sec}[1]{Section~\ref{#1}}

\newcommand{\Fig}[1]{Fig.~\ref{#1}}
\newcommand{\beq}{\begin{equation}}
\newcommand{\eeq}{\end{equation}}

\usepackage[T1]{fontenc}
\usepackage{ae,aecompl}

\usepackage{newtxtext,newtxmath}
\usepackage{natbib}
\bibpunct{(}{)}{;}{a}{}{,} 

\begin{document}

   \titlerunning{Massive dark matter-deficient satellite galaxies in cosmological simulations}
   \title{The Three Hundred: The existence of massive dark matter-deficient satellite galaxies in cosmological simulations}

   \author{A. Contreras-Santos
          \inst{1}
          \and
          F. Buitrago \inst{2,3}
          \and
          A. Knebe \inst{1,4,5}
          \and
          E. Rasia \inst{6,7}
          \and
          F. R. Pearce \inst{8}
          \and
          W. Cui \inst{1,4,9}
          \and
          C. Power \inst{5}
          \and 
          J. Winstanley \inst{5}
          }

   \institute{Departamento de F\'isica Te\'{o}rica, M\'{o}dulo 15, Facultad de Ciencias, Universidad Aut\'{o}noma de Madrid, 28049 Madrid, Spain \\
              \email{ana.contreras@uam.es}
         \and
             Departamento de F\'{i}sica Te\'{o}rica, At\'{o}mica y \'{O}ptica, Universidad de Valladolid, 47011 Valladolid, Spain
         \and
             Instituto de Astrof\'{i}sica e Ci\^{e}ncias do Espa\c{c}o, Universidade de Lisboa, OAL, Tapada da Ajuda, PT1349-018 Lisbon, Portugal
         \and
             Centro de Investigaci\'{o}n Avanzada en F\'isica Fundamental (CIAFF), Facultad de Ciencias, Universidad Aut\'{o}noma de Madrid, 28049 Madrid, Spain
         \and
             International Centre for Radio Astronomy Research, University of Western Australia, 35 Stirling Highway, Crawley, Western Australia 6009, Australia
         \and
             INAF – Osservatorio Astronomico di Trieste, via Tiepolo 11, I34131 Trieste, Italy
         \and
             IFPU – Institute for Fundamental Physics of the Universe, via Beirut 2, 34151, Trieste, Italy
         \and
             School of Physics \& Astronomy, University of Nottingham, Nottingham NG7 2RD, UK
         \and
             Institute for Astronomy, University of Edinburgh, Royal Observatory, Blackford Hill, Edinburgh EH9 3HJ, UK
             }

\date{Received January 1, 2020; accepted January 1, 2021}

 
  \abstract{
  The observation of a massive galaxy with an extremely low dark matter content (i.e. NGC 1277) has posed questions about how such objects form and evolve in a hierarchical universe. We here report on the finding of several massive, dark matter-deficient galaxies in a set of 324 galaxy clusters theoretically modelled by means of full-physics hydrodynamical simulations. We first focus on two example galaxies selected amongst the most massive and dark matter-deficient ones. By tracing the evolution of these galaxies, we find that their lack of dark matter is a result of multiple pericentre passages. While orbiting their host halo, tidal interactions gradually strip away dark matter while preserving the stellar component. 
  A statistical analysis of all massive satellite galaxies in the simulated clusters shows that the stellar-to-total mass ratio today is strongly influenced by the number of orbits and the distance at pericentres. Galaxies with more orbits and closer pericentres are more dark matter-deficient. Additionally, we find that massive, dark matter-deficient galaxies at the present day are either the remnants of very massive galaxies at infall or former central galaxies of infalling groups. 
  We conclude that such massive yet dark matter-deficient galaxies exist and are natural by-products of typical cluster galaxy evolution, with no specific requirement for an exotic formation scenario.}


   \maketitle
%

\section{Introduction}

In our current $\Lambda$CDM paradigm of galaxy formation and evolution, galaxies are thought to be formed within dark matter halos \citep{White1978}. Nevertheless, this paradigm still allows for the existence of dark matter-deficient galaxies. The so-called tidal dwarf galaxies \citep[TDGs,][and references therein]{Duc04}, are formed from tidal debris produced during interactions between galaxies, such as mergers or close encounters. Tidal forces disrupt and compress the gas and dust, triggering star formation and leading to the formation of these galaxies. TDGs generally have low masses and, since they retain little dark matter from their parents' halos, they are expected to be dark matter-deficient \citep{Barnes-Hernquist1992,Ploeckinger2018}. This formation mechanism is expected in a hierarchical context and, thus, their existence is not at variance with theoretical predictions from $\Lambda$CDM.

However, there have been many claims over the recent years about galaxies without dark matter in principle not associated with TDGs. \citet{VanDokkum18} reported that the dwarf galaxy NGC 1052-DF2 (also known simply as DF2) lacks dark matter by a factor of several hundred compared to the predictions of the standard galaxy formation and evolution model. Soon after that, \citet{VanDokkum19} announced the discovery of a second dark matter-deficient galaxy in the same group, NGC 1052-DF4 (or DF4). 
These publications have caused large controversy due to the way the dark matter mass was calculated, firstly because of the distance estimation to these objects \citep{Trujillo19,Monelli19} but also owing to the evidence for rotation in the case of DF2 \citep{Lewis20,Montes21}, implying that an alternative method for calculating the dynamical mass is needed. In contrast, other recent publications continue supporting the dark matter-deficiency of these objects \citep{Danieli2020,Shen2023}.

Assuming they are indeed dark matter-deficient, two main scenarios have been proposed for the formation of these dwarf galaxies. 
The first one involves a high-velocity collision of gas-rich galaxies very early on. This can be seen as a `mini Bullet cluster event', where the dark matter components do not interact and separate from their baryonic counterpart, thus producing a dark matter-deficient remnant. This scenario was proposed by \citet{Silk2019} and successfully reproduced in numerical simulations by \citet[see also \citealp{Lee2021}]{Shin2020}.

The second proposed scenario \citep{Ogiya2018,Nusser2020} is that these galaxies were formed as normal galaxies, but their dark matter halos were strongly stripped by tidal interactions, eventually making them appear as dark matter-deficient galaxies. Indeed, for satellite galaxies entering a host halo, interactions have been long known to be important regarding mass loss (see e.g. \citealp{Knebe2006} for a quantification of this effect in N-body simulations, or \citealp{Joshi2019} for a more recent study with hydrodynamics simulations). 
Different numerical simulations have demonstrated that, under certain conditions, dark matter-deficient galaxies can be produced following this mechanism without contradicting $\Lambda$CDM predictions \citep{Ogiya2018,Jing2019,Maccio2021,Jackson2021,Ogiya22}. Observations showing that DF4 presents tidal tails have supported this scenario \citep{Montes20,Keim2022,Golini2024}, while for DF2 the situation is still unclear, with some studies claiming the presence of tidal features \citep{Keim2022}, and others stating that, even with very deep images, there is no sign of tidal distortion \citep{Golini2024}.


While these dark matter-deficient galaxies are all low-mass galaxies (for instance, DF2 has a stellar mass of $M_* \sim 2 \times 10^8$ M$_\odot$), a recent publication \citep{Comeron2023} has claimed that NGC 1277, a massive galaxy (stellar mass M$_{*}$ $\sim 1.6 \times 10^{11}$ M$_{\odot}$) located in the Perseus galaxy cluster, is compatible with lacking dark matter, too. There are claims that this is at odds with theoretical predictions based on $\Lambda$CDM \citep{Santucci22}. Although similar mechanisms as the ones described before can be invoked for the formation of such a galaxy, the high stellar mass of NGC 1277 has included a new feature to check for and thus added a degree of complexity to this issue. 

While dark matter-deficient dwarf galaxies have been observed before and are possibly explained by the previously mentioned scenarios, for massive dark matter-deficient galaxies there are apparently only two works investigating them using cosmological simulations, that is, \citet[][using the Illustris simulations]{Yu2018} and \citet[][using both Illustris and EAGLE simulations]{Saulder2020}. But the work of \citet{Saulder2020} has clearly demonstrated that the appearance of these massive dark matter-deficient galaxies -- as reported in \citet{Yu2018} -- is related to non-physical events happening during the crossing of the periodic boundaries of the simulations used in those studies. As it stands there appears to be no conclusive work that provides evidence of the existence or even a credible formation scenario for massive dark matter-deficient galaxies.

Motivated by this situation, in this work we explore the 324 theoretically modelled galaxy clusters of \textsc{The Three Hundred}\footnote{\url{http://www.the300-project.org}} project \citep{Cui2018} in order to test whether there is any physical phenomenon able to occasionally produce massive dark matter-deficient galaxies within galaxy clusters. In Section \ref{sec:data} we introduce the simulations' data set used, while in Section \ref{sec:dmd} we present the massive dark matter-deficient galaxies that we find within these clusters and the formation mechanism we propose for them. We finish by summarising and discussing the results in Section \ref{sec:conclusions}.


\section{Data and methodology} \label{sec:data}

\subsection{The Three Hundred sample}

The data used in this work comes from \textsc{The Three Hundred} project, which is a suite of simulations of 324 large galaxy clusters modelled by full-physics hydrodynamical re-simulations. For details of the simulations we refer the reader to the numerous studies based upon it, first and foremost \cite{Cui2018} where the project has been introduced. We only briefly summarise here the most important aspects.

\textsc{The Three Hundred} data set consists of a suite of 324 spherical regions of diameter $30 \hMpc$ centred on the most massive objects found within a parent cosmological dark matter only simulation of side length $1 \hGpc$ (MultiDark Planck 2, MDPL2, \citealp{Klypin2016}). After splitting the initial dark matter particles into dark matter and gas, those regions have been re-simulated from their initial conditions including now full hydrodynamics using the smoothed particle hydrodynamics (SPH) code \textsc{Gadget-X}.
This is a modified version of the non-public \textsc{Gadget3} code\footnote{See \citet{Springel2005} for a detailed description of the last public version \textsc{Gadget-2}} \citep{Murante2010,Rasia2015,Planelles2017,Biffi2017}, which features an improved SPH scheme \citep{Beck2016} that gives a better description of discontinuities and instabilities within the gas. Star formation is carried out as in \citet{Tornatore2007} and follows the star formation algorithm presented in \citet{Springel2003}, while black hole growth and AGN feedback are implemented following \citet{Steinborn2015}.

For this work we focus on the 324 clusters that lie on the centre of each region, as modelled by \textsc{Gadget-X}. The dark matter (gas) mass resolution is $12.7 \cdot 10^{8}h^{-1}M_\odot$ ($2.4 \cdot 10^{8}h^{-1}M_\odot$). Stellar particles have an average mass of $0.4\cdot 10^{8}h^{-1}M_\odot$ (as in \textsc{Gadget-X} a gas particle spawns multiple star particles). The spatial resolution is quantified by a Plummer equivalent softening of 6.5 $h^{-1}$kpc. A total of 128 different snapshots have been stored for each simulation from redshift $z = 17$ to $0$.




\subsection{Halo catalogues and merger trees}

\begin{figure*}
\centering
  \includegraphics[width=15cm]{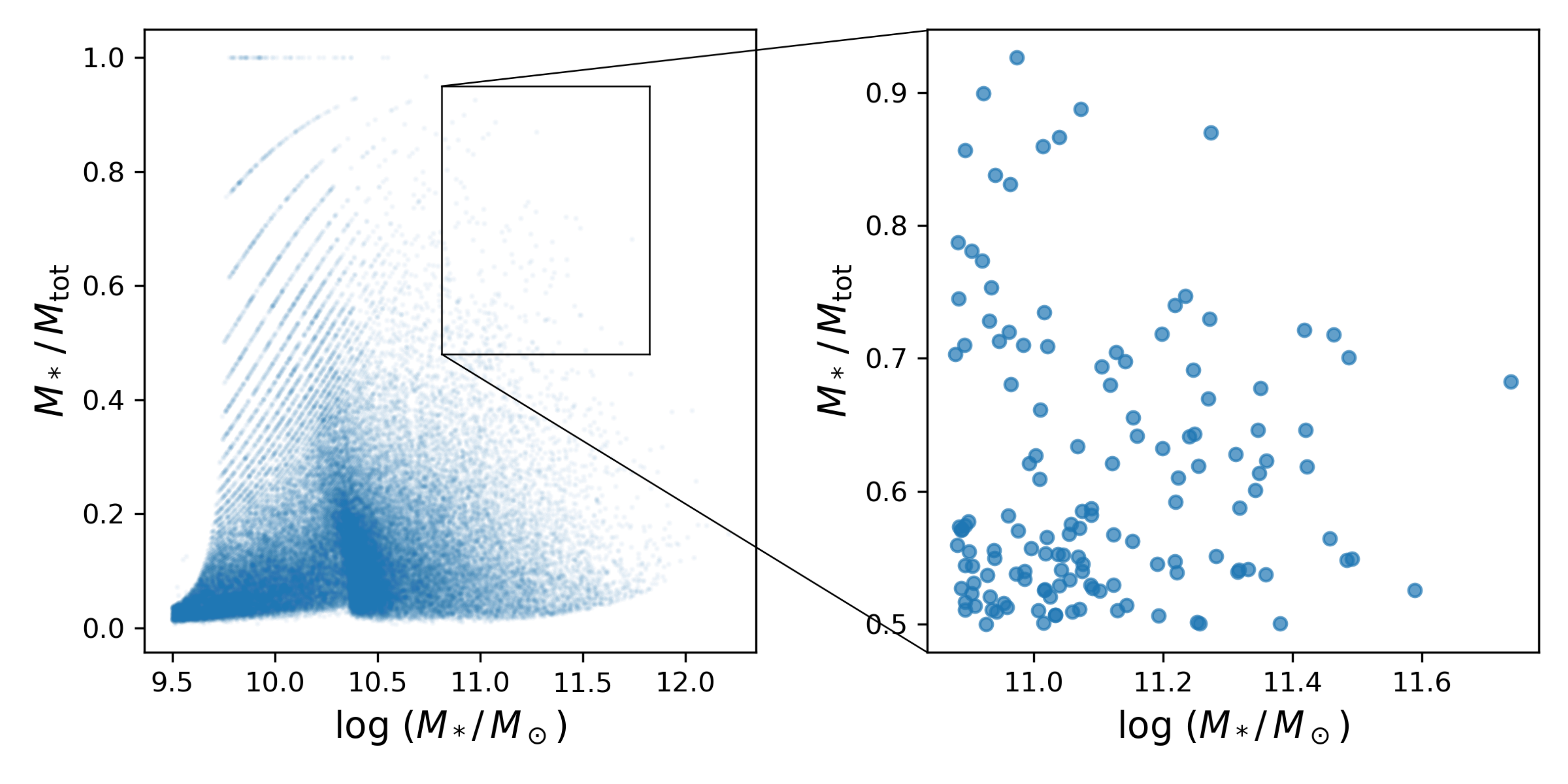}
  \caption{Stellar-to-total mass ratio as a function of stellar mass. Left, for all the selected galaxies at $z=0$. Right, zoom-in to the most massive and dark matter-deficient galaxies.}
  \label{fig:ratio-mstar}
\end{figure*}

The halo and galaxy catalogues, respectively, have been generated by the open-source object finder for cosmological simulations \textsc{AHF}\footnote{\url{http://popia.ft.uam.es/AHF}} \citep{KnollmannKnebe2009, Gill2004}. Halo masses and edges are defined as over-densities $200 \times \rho_{\rm crit}$ and hence denoted as $M_{200}$ and $R_{200}$, respectively. We note that for subhalos -- due to their embedding within the host halo's density field -- the masses are actually given by the mass gravitationally bound to the subhalo's centre, denoted as $M_{\rm tot}$ \citep[see the Appendix of][for more details]{KnollmannKnebe2009}. 

We like to remark that the \textsc{AHF} halo finder takes all particles into account when searching for objects, which allows us to identify star-only (or even gas-only) objects in simulations, that is, the objects of interest for this study. Note that other finders start with constructing dark matter-only halos, subsequently adding star and gas particles to it \citep[cf.][for \textsc{SUBFIND}]{Knebe2013,Dolag2009}, making the identification of dark matter-deficient objects intrinsically difficult.

To track the evolution of identified structures across different simulation snapshots, we utilised the \textsc{MergerTree} code that comes with the halo finder, adapted specifically to trace stellar particles in addition to dark matter. This -- novel -- adaptation is an important feature for our study, as we are primarily interested in dark matter-deficient (if not dark matter-free) objects, where it is essential to follow the stellar component's history across snapshots. We further note that the latest version of \textsc{MergerTree} applied here also allows for `snapshot skipping', which is an essential feature that post-corrects for halo finder incompleteness \citep{Srisawat2013,Wang2016}.


\section{Dark matter-deficient galaxies} \label{sec:dmd}

\subsection{Finding dark matter-deficient galaxies}

The first question we want to answer is whether massive dark matter-deficient satellite galaxies exist in our simulated clusters. That is, our goal is to find galaxies that are dark matter-deficient and, at the same time, have a high stellar mass. We first select all the satellite galaxies within the simulations by defining our initial sample, from which we will make further sub-selections. 
For this initial selection we restrict our objects to comply with the following criteria at redshift $z=0$: distance less than $R_{200}$ to the host cluster centre, stellar mass $M_* \geq 10^{9.5} M_\odot$ and total mass $M_\mathrm{tot} < 10^{13} M_\odot$.\footnote{$M_{\rm tot}$ is the sum of dark matter, gas, and stars.} The latter ensures that we do not include the brightest cluster galaxy in our study. 

In \Fig{fig:ratio-mstar} we show the ratio of stellar-to-total mass, $M_*/M_\mathrm{tot}$, as a function of stellar mass $M_*$ for our initial sample (left panel). Before going into more detail, there are two distinct features to note here. First, the `stripes' in the upper left region of the plot are due to limited mass resolution: the horizontal line in the upper left corner is caused by galaxies with no dark matter (DM) particles ($M_* = M_\mathrm{tot}$), while the curves below are for galaxies with only one, two, three, etc. DM particles. The second feature is regarding the values of the stellar masses of the galaxies in \textsc{Gadget-X}, i.e. the $x$-axis of the plot, in which the density of points indicates that there is a peak at masses $\log M_*/M_\odot \gtrsim 10.5$. This was already observed and discussed in \citet{Cui2018} and \citet{Cui2022}. Figs. 7 and 8 in these works, respectively, already show this peak in the satellite galaxies stellar mass function. As shown in \Fig{fig:ratio-mstar}, neither will affect our results given the mass range of the galaxies we select ($\gtrsim 10^{11} M_\odot$, see below).

Focusing now on the $y$-axis values, we can see in the left panel of \Fig{fig:ratio-mstar} that the great majority of the galaxies have a stellar-to-total mass ratio $M_*/M_\mathrm{tot} \lesssim 0.2$ at $z=0$. Since we are interested in massive, dark matter-deficient galaxies, we focus on the upper right corner of this plot. The right panel of \Fig{fig:ratio-mstar} shows a zoom-in to this region, where all the galaxies have $M_*/M_\mathrm{tot} \geq 0.5$. This indicates that -- across all 324 (stacked) host clusters -- there is quite a number of dark matter-deficient galaxies, which have, at least, the same mass in stars as in dark matter plus gas. Furthermore, all the galaxies in the right panel of \Fig{fig:ratio-mstar} have a stellar mass $M_* \geq 7.5 \cdot 10^{10} M_\odot$, and thus can be considered massive dark matter-deficient galaxies. We select this threshold as it implies that all the objects in the right panel of \Fig{fig:ratio-mstar} contain more than 1000 (star) particles today and hence were even more massive in the past, including their infall time. They are therefore well enough resolved according to recent convergence studies \citep[e.g.][]{Allgood2006,vandenBosch2018a,Green2021,Errani2021}. \Fig{fig:ratio-mstar} illustrates the existence of massive galaxies with very low dark matter content (reaching even $M_*/M_\mathrm{tot} \gtrsim 0.8$) in cosmological simulations. We will devote the next subsections to understanding their formation.



\subsection{Example case studies} \label{sec:DMD-examples}

\begin{figure*}
\centering
  \includegraphics[width=18.3cm]{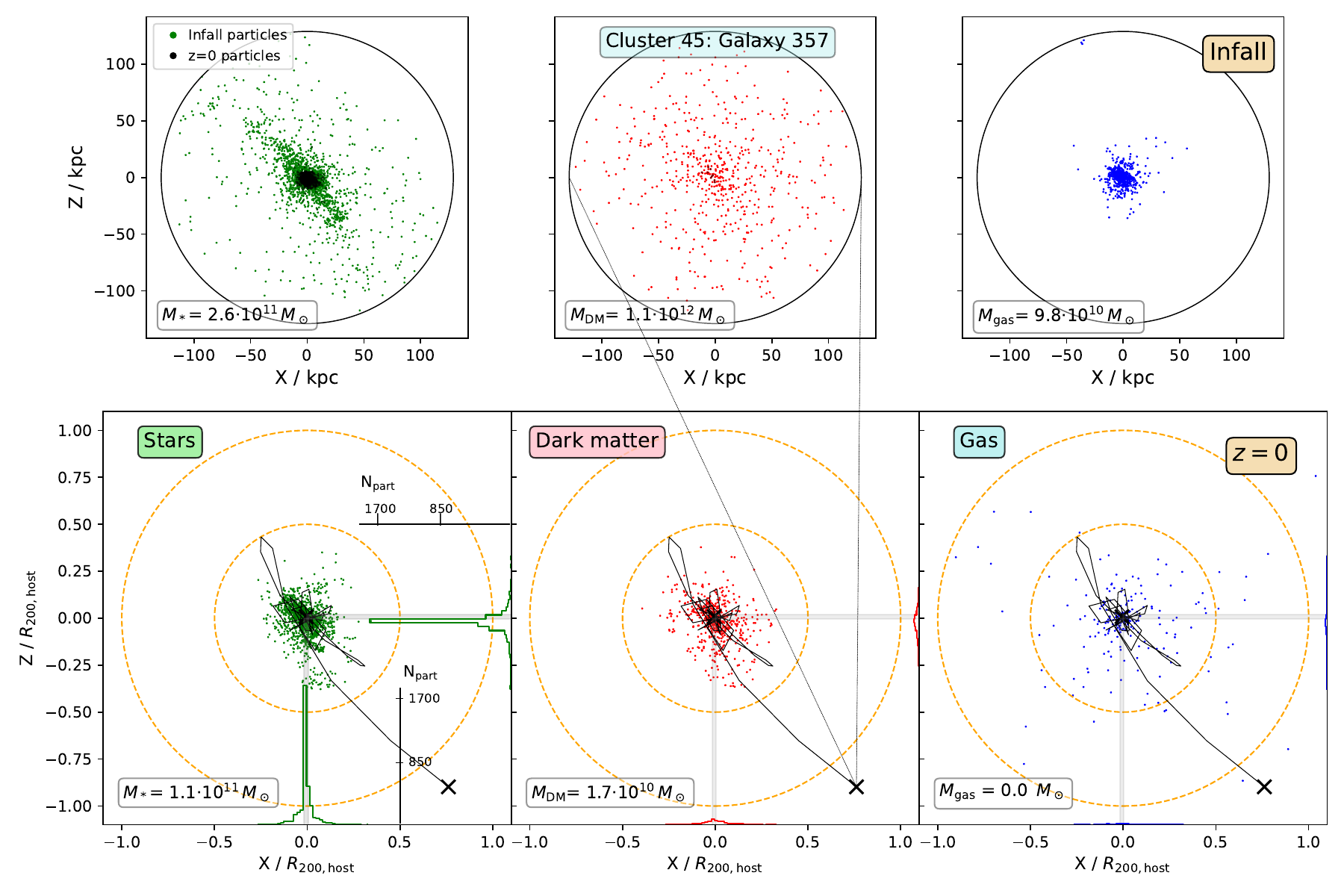}
  \caption{For one of the most massive and dark matter-deficient galaxies, we display in the first row the galaxy at infall, separating into, from left to right, stellar (green), dark matter (red), and gas (blue) particles. The coloured particles are the ones that belong to the object at infall time, while in black we mark those that also end up belonging to the object at $z=0$. In the second row we show the final position of all the particles above, now in units of the host halo radius $R_{200}$. The histograms in the bottom and right axes of each plot show the distribution of the positions of the particles, where the shaded grey regions correspond to the radius of the galaxy at $z=0$. The trajectory of the galaxy from infall until $z=0$ is shown as a black solid line. The two circles correspond to the host radius $R_{200}$ at $z=0$ (outer circle) and $0.5 \cdot R_{200}$ (inner circle).}
  \label{fig:evolution-particles1}
\end{figure*}

\begin{figure*}
\centering
  \includegraphics[width=18.3cm]{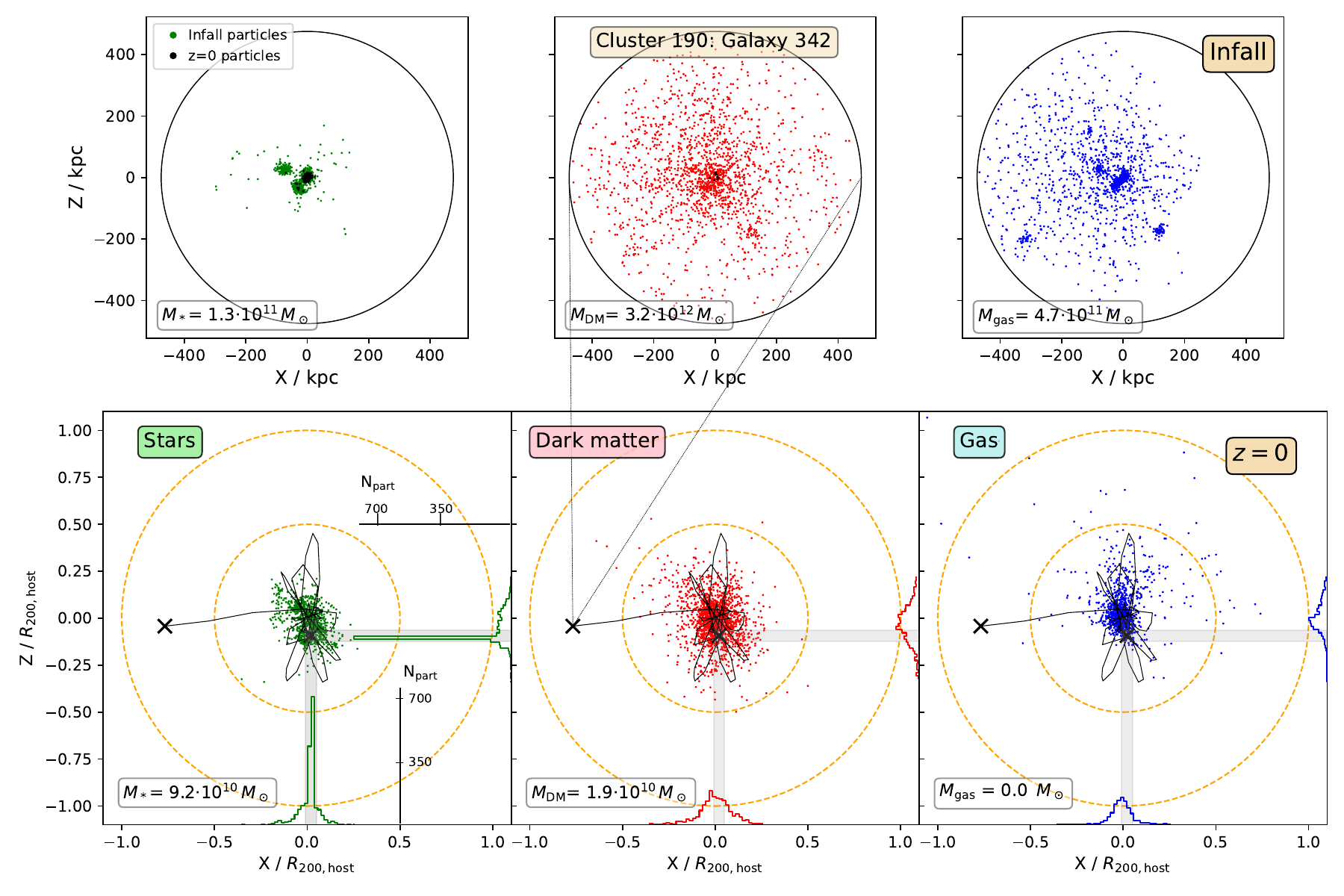}
  \caption{Same as \Fig{fig:evolution-particles1} but for another object selected from the most extreme ones in the right panel of \Fig{fig:ratio-mstar}.}
  \label{fig:evolution-particles2}
\end{figure*}

Having identified the most extreme cases of massive dark matter-deficient galaxies in our cluster simulations, the goal now is to understand their formation and the reason for their lack of dark matter.
For this purpose, we follow the trajectories of these objects back in time to investigate their evolution. In this section, we select two representative examples of the most extreme objects in \Fig{fig:ratio-mstar} and show how their properties have changed since they were outside the cluster until the present day. 

In \Fig{fig:evolution-particles1} we focus on one galaxy that, at $z=0$, has $M_* = 1.1 \cdot 10^{11} M_\odot$ and $M_*/M_\mathrm{tot} = 0.87$. The `names' of the host cluster and subhalo are indicated in the text box at the top. We first show, in the upper row, the galaxy at infall time, that is, in the last snapshot where the distance $r$ to the centre of the host cluster was $r > R_{200}$. From left to right, the dots show the positions of the stellar (green), dark matter (red) and gas (blue) particles that belonged to the galaxy at this time. In each panel we also indicate in black the particles that remain in the object at $z=0$. The mass of each of the components is also indicated in the lower left corner. It can be seen that, at infall, the object was a massive galaxy surrounded by an extended dark matter halo, as expected for a galaxy of this kind. In the lower row we show the positions of these same particles at $z=0$, now in the reference frame of the host cluster. The two dashed circles correspond to the host radius $R_{200}$ and $0.5 \cdot R_{200}$, while the trajectory of the galaxy from infall until $z=0$ is depicted as a black solid line. This line shows the different orbits conducted around the cluster centre, where it can be seen how all the orbits are within $0.5 \cdot R_{200}$. Focusing on the positions of the particles, we see that the same particles are now spread over a much larger area. For better visualisation of the situation, we have included at the bottom and right axes of each plot a histogram with the distributions of the particles along each axis. The grey shaded regions correspond to the radius\footnote{Here the radius is defined as the distance to the last bound particle.} of the galaxy at $z=0$, in order to give an approximated idea of which particles still belong to the object. The mass can again be seen in the lower left corner of each panel. Here it is clear that, while stars are being stripped (the stellar mass is reduced by a factor $\sim 2$ from infall to $z=0$), the stellar particles remain much more concentrated in the centre of the galaxy even after all the orbits. The black dots in the upper row confirm that the particles that remain bound to the galaxy are those that were closer to the centre at the beginning, as expected. The dark matter and gas particles, on the contrary, are much more spread at $z=0$, and their masses have been much more dramatically reduced (two orders of magnitude in the case of the dark matter, and completely stripped for the gas).

A second example of a massive dark matter-deficient galaxy is shown in \Fig{fig:evolution-particles2}. In this case, the properties at $z=0$ are $M_* = 9.2 \cdot 10^{10} M_\odot$ and $M_*/M_\mathrm{tot} = 0.83$. Similarly to \Fig{fig:evolution-particles1}, in the upper row of \Fig{fig:evolution-particles2} we depict the object at infall time. In contrast to the previous example, we now see that what we have is a group of galaxies, in which three different blobs can be distinguished. They share a common dark matter halo, whose mass is also within the expected range for this kind of object. The stellar particles are much more concentrated towards the centre than the dark matter and the gas. We also note the different scales in the upper row axes between \Fig{fig:evolution-particles1} and \Fig{fig:evolution-particles2}, indicating the different sizes of the objects at infall. In the bottom row of \Fig{fig:evolution-particles2} we see that, despite the slightly different initial setup, the situation is very similar to the previous example (\Fig{fig:evolution-particles1}). The group conducted many orbits around the cluster centre, in which the gas and the dark matter were significantly stripped, with the remaining particles being spread around the inner part of the cluster. The remaining stellar particles, on the contrary, show a distribution that sharply peaks at the final position of the object. The black dots in the upper row reveal that only one of the initial blobs remains at $z=0$, while the rest were stripped and account for the stellar mass lost, leading to the resulting massive dark matter-deficient galaxy. 

We want to highlight that, after investigating the properties at both infall time and present day of the most extreme objects in the right panel of \Fig{fig:ratio-mstar}, we find that these two examples are representative of the whole population. The resulting massive dark matter-deficient galaxies are the result of either a massive galaxy at infall that loses its dark matter halo, or an infalling group with an extended dark matter halo, in which the central galaxy remains in the centre after infall and the dark matter halo is heavily stripped.

\begin{figure}
\centering
  \includegraphics[width=8.5cm]{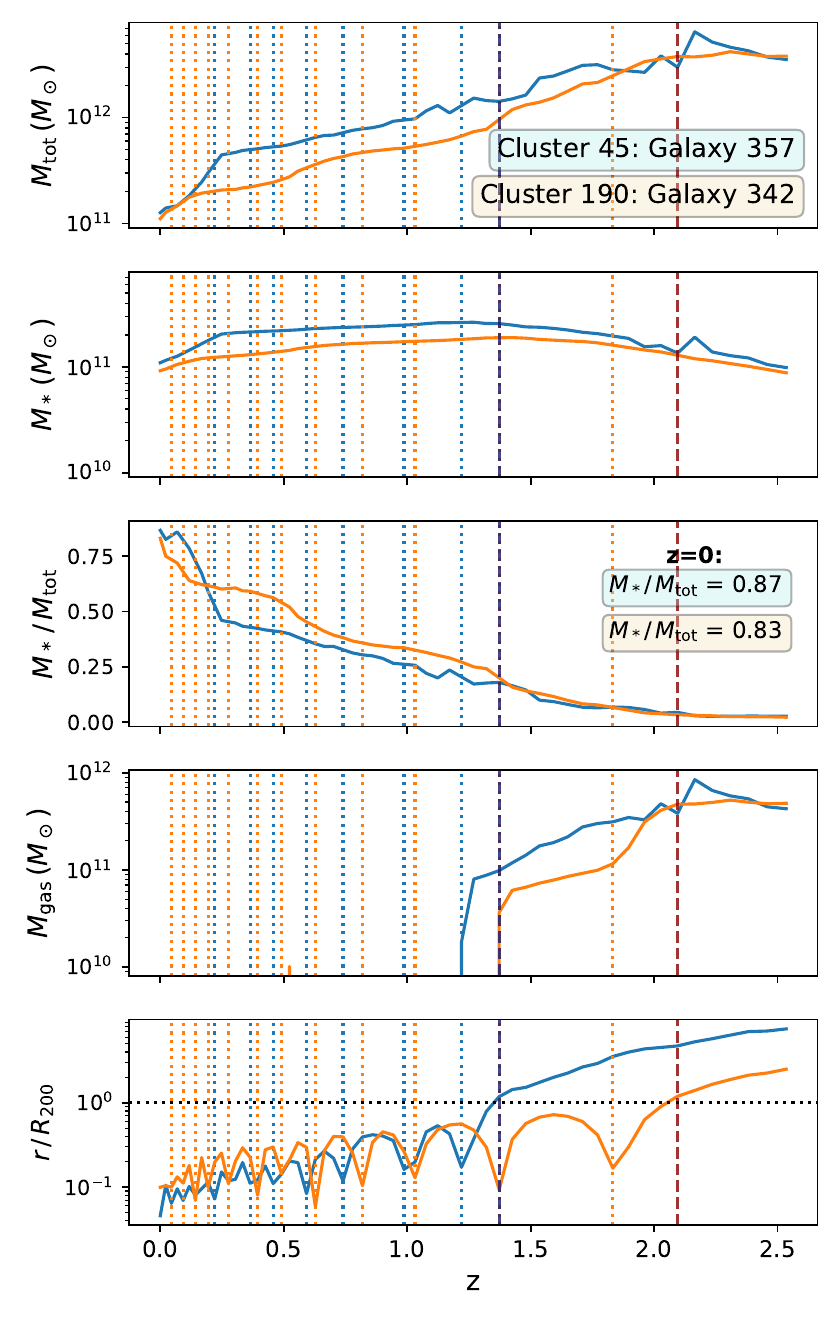}
  \caption{For the two objects previously selected as examples (Figs.~\ref{fig:evolution-particles1} and \ref{fig:evolution-particles2}), we display the evolution with redshift of five different quantities. From top to bottom, total mass $M_\mathrm{tot}$, stellar mass $M_*$, stellar-to-total mass ratio $M_*/M_{\rm tot}$, gas mass $M_\mathrm{gas}$ and distance to cluster centre in units of cluster radius, $r/R_{200}$. The vertical blue and orange dotted lines indicate minima in the distance to centre after infall, i.e. pericentres of the orbit, which can be seen correspond to minima in $M_\mathrm{tot}$ and maxima in $M_*/M_\mathrm{tot}$. The stellar mass and stellar-to-total mass ratio of the object at $z=0$ are indicated in the fourth panel. The horizontal dotted line in the bottom panel indicates the distance $r = R_{200}$, i.e., pointing to the infall to the cluster. The infall redshift of both objects is indicated in each plot with a dark vertical dashed line.}
  \label{fig:evolution-rdist}
\end{figure}

To complement the study in Figs.~\ref{fig:evolution-particles1} and \ref{fig:evolution-particles2}, we now investigate in more detail how the masses of the objects evolve in relation to orbital parameters. For the same two examples, we now show in \Fig{fig:evolution-rdist} (from top to bottom, solid lines) the total mass $M_\mathrm{tot}$, the stellar mass $M_*$, the stellar-to-total mass ratio $M_*/M_\mathrm{tot}$, the gas mass $M_\mathrm{gas}$, and the distance to cluster centre in units of the cluster radius $r/R_{200}$ (in logarithmic scale) as a function of redshift $z$. For reference, the stellar-to-total mass ratios of the galaxies at $z=0$ are indicated in the third panel, as are the `names' of the host cluster and subhalo. We show at the same time the object in \Fig{fig:evolution-particles1} (blue lines, galaxy with ID 357 in host cluster 45), and that in \Fig{fig:evolution-particles2} (orange lines, infalling group with ID 342 in host cluster 190). The vertical dotted lines represent the pericentre passages as derived from the bottom panel, while the dashed lines indicate the infall times of the two galaxies (in dark blue and dark orange, respectively). The horizontal dotted line marks the crossing of the host's $R_{200}$.  We also like to remark that the first two panels, for $M_\mathrm{tot}$ and $M_*$, represent different orders of magnitude in the $y$-axis, but the dynamical range shown is the same for both, so that the relative changes of these masses are equivalent in both plots. 

In both cases, we clearly see the gradual decrease in total mass after the galaxy enters the host's $R_{200}$. At the same time, the stellar-to-total mass ratio increases, while the stellar mass remains relatively constant during this time. This pattern, which shows a noticeable uptick during pericentre passages (where the slopes become steeper), suggests that the DM component of the galaxy is being stripped after entering the host environment, with an enhanced effect when passing close to the cluster centre. The stellar mass, however, is being held by the galaxy, making the stellar-to-total mass ratio grow and reach the present values. The gas has already been completely removed after the second pericentre passage. \Fig{fig:evolution-rdist} further shows how the galaxy gradually spirals towards the centre\footnote{We have checked that this is a real effect and not due to an increase of the host's radius, i.e. not normalising by the host's radius shows the same shrinking, indicating that the main driver of this shrinking is the orbit and not the cluster's size increase since $z\sim 2.5$. See Appendix \ref{appendix:host-radius} for more details.}. We note that, due to the cadence of our simulation outputs, the pericentre values shown here are simply upper bounds and could in fact be much smaller. Although a correction such as the one promoted by, for instance, \citet{Richards2020} could be applied to the orbits, we believe that these upper bounds are sufficient for the results presented here.


This scenario where subhalos lose dark matter mass due to their interaction with their host has been widely studied in previous works by means of numerical simulations \citep[see e.g.][]{Mayer2002,Hayashi2003,Gill2004,Kravtsov2004,Penarrubia2008,Ogiya2018,Errani2020,Nusser2020,Green2021,Stuecker2023}. This stripping is produced by tidal forces, that act first on the outer DM halo, while the stellar component, which is much more centrally concentrated and protected by the DM halo, is harder to strip \citep{Choi2009,Libeskind2011,Smith2016}. More recent works have also studied in detail the process of satellite mass loss, constructing models that can predict the efficiency of these tidal forces in stripping the subhalos \citep{Drakos2020,Errani2021}. Using the Illustris simulation, \citet{Niemiec2019} found that subhalos start losing DM when they get closer than $1.5 R_\mathrm{vir}$ of their host centre, and that this mass loss during infall deeply affects their stellar-to-halo mass ratio. We further remark that the \textsc{Gadget-X} clusters host rather massive central BCGs \citep[see, for instance, Fig.8 in][]{Contreras-Santos2022a} that eventually induce enhanced stripping. In this work we show that the commonly accepted framework of tidal stripping can lead to massive dark matter-deficient galaxies. In this subsection we have used two different examples to illustrate this mechanism. In the following subsection we will select a wider sample of massive galaxies and statistically discuss the specific conditions required for the formation of these objects based upon this mechanism and our particular \textsc{The Three Hundred} simulations performed with \textsc{Gadget-X}.



\subsection{Statistical study} \label{sec:dmd-statistical}

While \Fig{fig:evolution-rdist} clearly shows how tidal interactions can strip off nearly all dark matter (yet leaving the stellar component practically untouched), the question remains why some particular subhalos are losing much more DM than other cluster satellite galaxies, which also orbit the cluster and experience comparable tidal forces. To investigate this, we now select a sample of massive galaxies and statistically examine the relationship between their stellar-to-total mass ratio and other galaxy properties. 

Starting with the initial sample of satellite galaxies from the left panel of \Fig{fig:ratio-mstar} (i.e., those inside $R_{200}$ of their host cluster at $z=0$ and with $M_\mathrm{tot} < 10^{13} M_\odot$), we further refine our selection to include only those with stellar masses $M_* \geq 7.5 \cdot 10^{10} M_\odot$. By studying the stellar-to-total mass ratio of these massive galaxies today as a function of different properties, we aim to understand the formation of massive dark matter-deficient galaxies compared to other galaxies that survive until $z=0$ (being still well-resolved) and retain a typical amount of dark matter.

\subsubsection{Number of orbits and pericentre distance}

\begin{figure}
\centering
  \includegraphics[width=9cm]{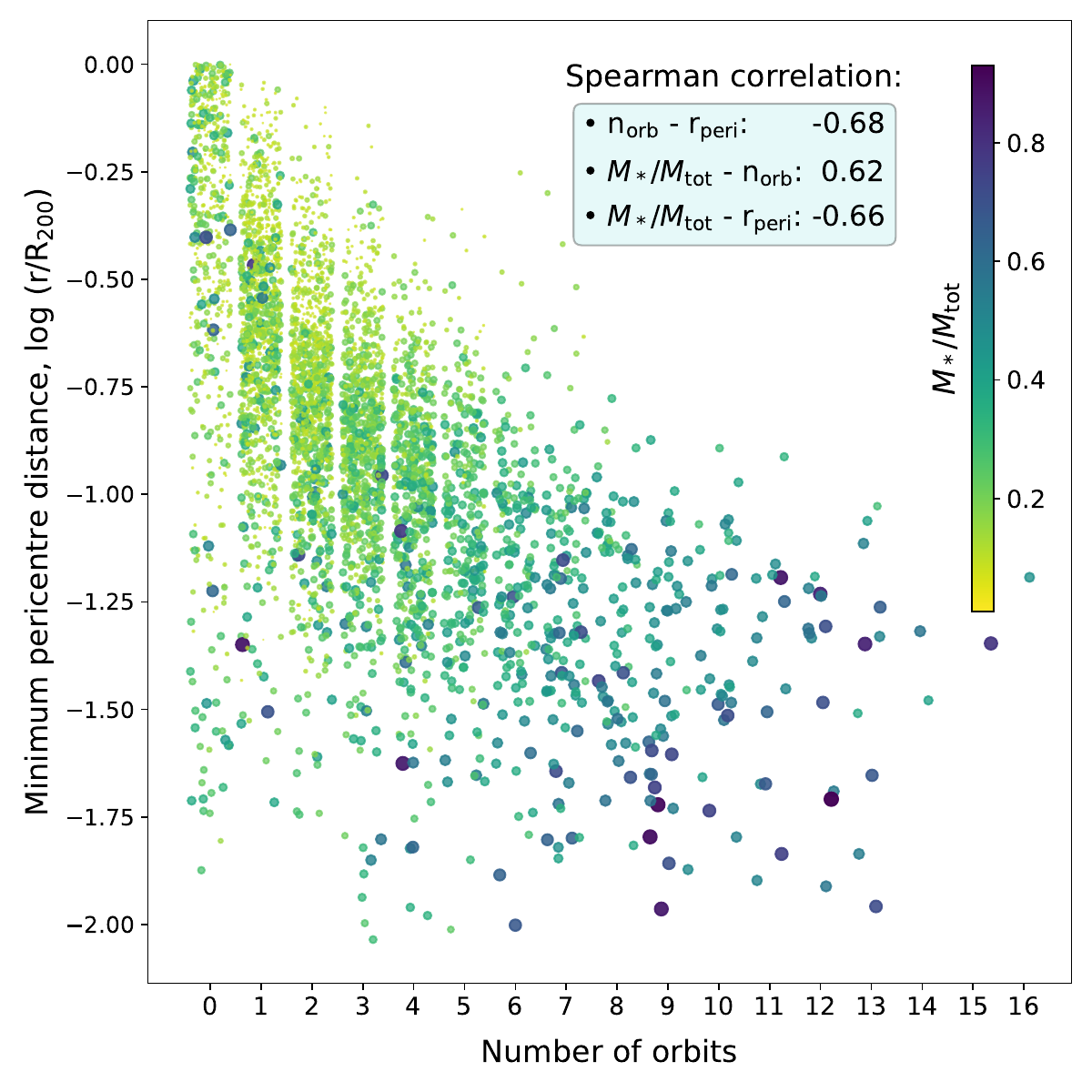}
  \caption{Minimum pericentre distance (in units of $R_{200}$ of the cluster) as a function of the number of orbits performed by each galaxy, for the sample of massive satellite galaxies. The colours (and size) indicate the stellar-to-total mass ratio, as seen in the central colourbar, with darker (and larger) dots depicting galaxies with a higher value of $M_*/M_\mathrm{tot}$, i.e., with less DM.}
  \label{fig:distance-orbits}
\end{figure}


In the bottom panel in \Fig{fig:evolution-rdist}, both example galaxies appear to complete many orbits around the host cluster, reaching very close to its centre. Motivated by this, we trace all the selected satellite galaxies back in time and we calculate the number of orbits inside the host cluster, as well as their distances at pericentres. These calculations are based on the minima in the distance from each galaxy to the cluster centre, as illustrated in \Fig{fig:evolution-rdist}.


In \Fig{fig:distance-orbits} we show the minimum pericentre distance (in units of the host's radius $R_{200}$) as a function of the number of orbits after infall. We use the minimum pericentre distance, instead of, for instance, the median across the different orbits, since the cadence of our simulations means that the measured pericentres are upper bounds. Hence, the minimum distance can be more representative of the actual orbit than other quantities such as the median pericentre distance. Nevertheless, we have checked that similar trends remain if using the median or the mean pericentre distance. For visualisation purposes, in \Fig{fig:distance-orbits} we randomly shift the dots on the $x$-axis by a small amount, so they do not fall all on the same line, but we remark that the number of orbits is computed as a discrete quantity. The colour scheme for the dots is indicative of the stellar-to-total mass ratio, as seen in the central colourbar, where darker means a higher ratio (i.e. lower dark matter fraction). To further focus on the most dark matter-deficient objects, we also make the size of the dots proportional to their stellar-to-total mass ratio, so that darker dots are larger too. 

The first apparent feature in \Fig{fig:distance-orbits} is the --expected-- correlation between pericentre and number of orbits. The Spearman correlation coefficient between these two variables is indicated in the text box. Besides, \Fig{fig:distance-orbits} clearly shows a strong dependence of the stellar-to-total mass ratio on both the number of orbits and the pericentre distance. In general, the most dark matter-deficient galaxies lie at the bottom right part of the plot, with exceptionally high number of orbits and minimum pericentres close to the host's centre. This is confirmed by the Spearman correlation coefficients, which indicate a strong positive correlation of $M_*/M_\mathrm{tot}$ with the number of orbits, and a similar but negative correlation with the minimum pericentre distance.

We also want to remark that -- according to \Fig{fig:distance-orbits} -- there are some galaxies in the sample with a significantly high stellar-to-total mass ratio, but with a lower number of orbits and high pericentre distance. They show up as blue dots that lie close to the upper left corner of the plot. Although we have not studied them in detail here, we suggest the possibility of them being `renegade' or `Hermeian' halos, which have changed their affiliation from one host to another. These objects have been studied within the context of the Local Group \citep{Knebe2011, Newton2022} and further confirmed by \citet{Osipova2023} to exist around almost every halo using large-scale structure cosmological simulations. 

\subsubsection{Infall time}

\begin{figure}
\centering
  \hspace*{-0.2cm}
  \includegraphics[width=9cm]{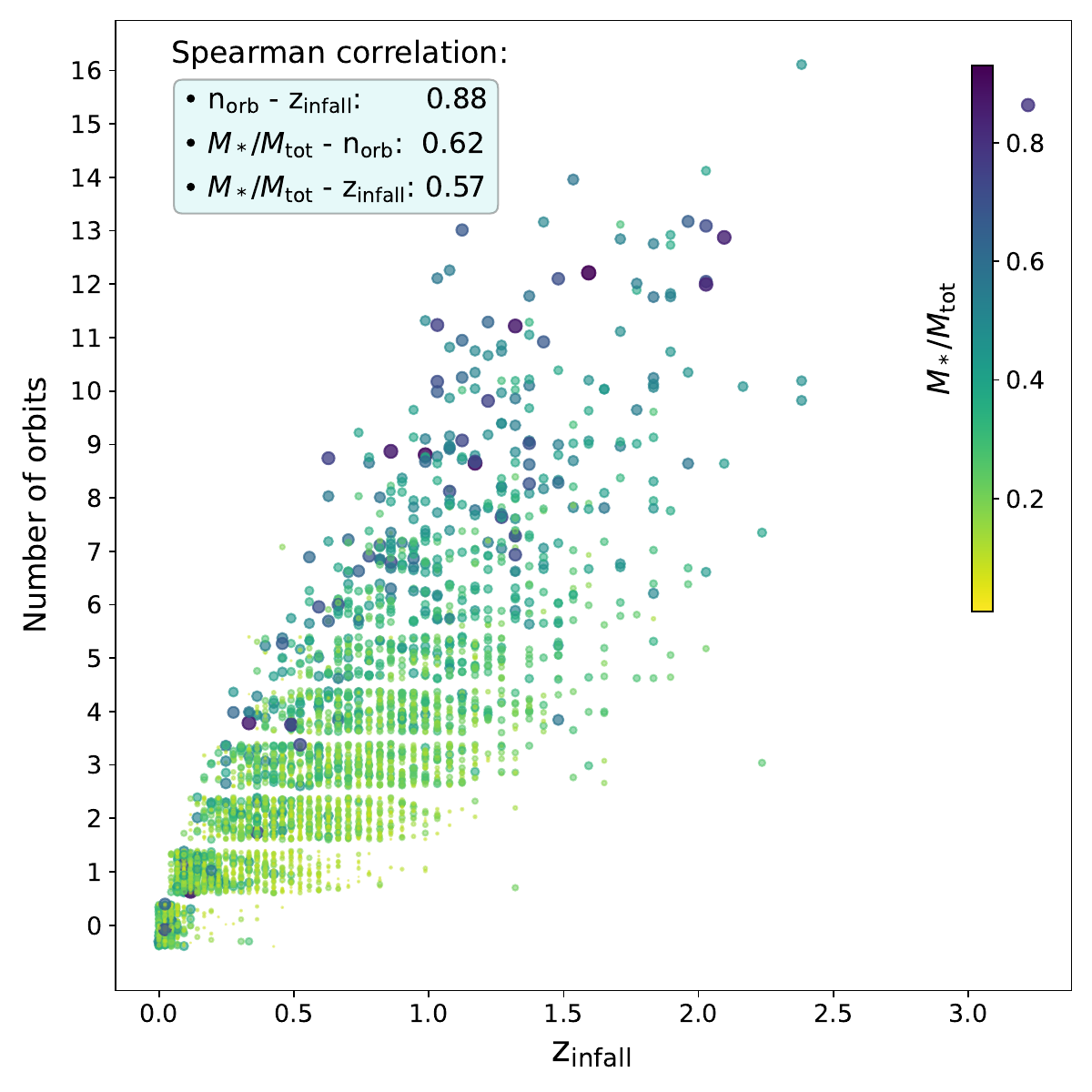}
  \caption{Number of orbits as a function of infall time for the sample of massive satellite galaxies. The colours (and size) indicate the stellar-to-total mass ratio, as seen in the central colourbar, with darker (and larger) dots depicting galaxies with a higher value of $M_*/M_\mathrm{tot}$, i.e., with less DM.}
  \label{fig:orbits-zinfall}
\end{figure}

We further analyse the evolution of the selected galaxies by focusing on the time they fall into the host cluster. We get the infall redshift of the galaxies as the last snapshot where $r/R_{200} > 1$. In \Fig{fig:orbits-zinfall} we show this value as a function of the number of orbits that each galaxy conducts around the cluster after infall. As in \Fig{fig:distance-orbits}, the $y$-axis values are slightly moved for better visualisation. The dots are again coloured by their stellar-to-total mass ratio at $z=0$, with their size also proportional to this value, in order to highlight the objects of interest, that is, the dark matter-deficient ones. We see again a positive correlation between $M_*/M_\mathrm{tot}$ and $z_\mathrm{infall}$, which means that dark matter-deficient galaxies tend to have higher infall redshifts. When correlating to number of orbits, we find -- as seen in the previous plot -- a clear correlation, but it is interesting to see that the purple dots (the most dark matter-deficient objects) lie systematically above the general relation for all the galaxies. This means that these objects had more orbits than a general object (in terms of dark matter content) that entered the host at about the same time. This could be due to the dark matter-deficient objects moving faster or because, being closer to the centre, their orbits are shorter. 

In line with these findings, \citet{Smith2022} found, using a set of dark matter-only simulations to investigate tidal stripping, that the strength of tidal mass loss is related to the orbital velocity of the infalling objects, as well as to their environment at infall time. We leave a more detailed investigation of this for the future, where we plan to connect the entry points of the galaxies to the surrounding cosmic web, too.


\subsubsection{Stellar mass at infall}

\begin{figure}
\centering
  \hspace*{-0.2cm}
  \includegraphics[width=9cm]{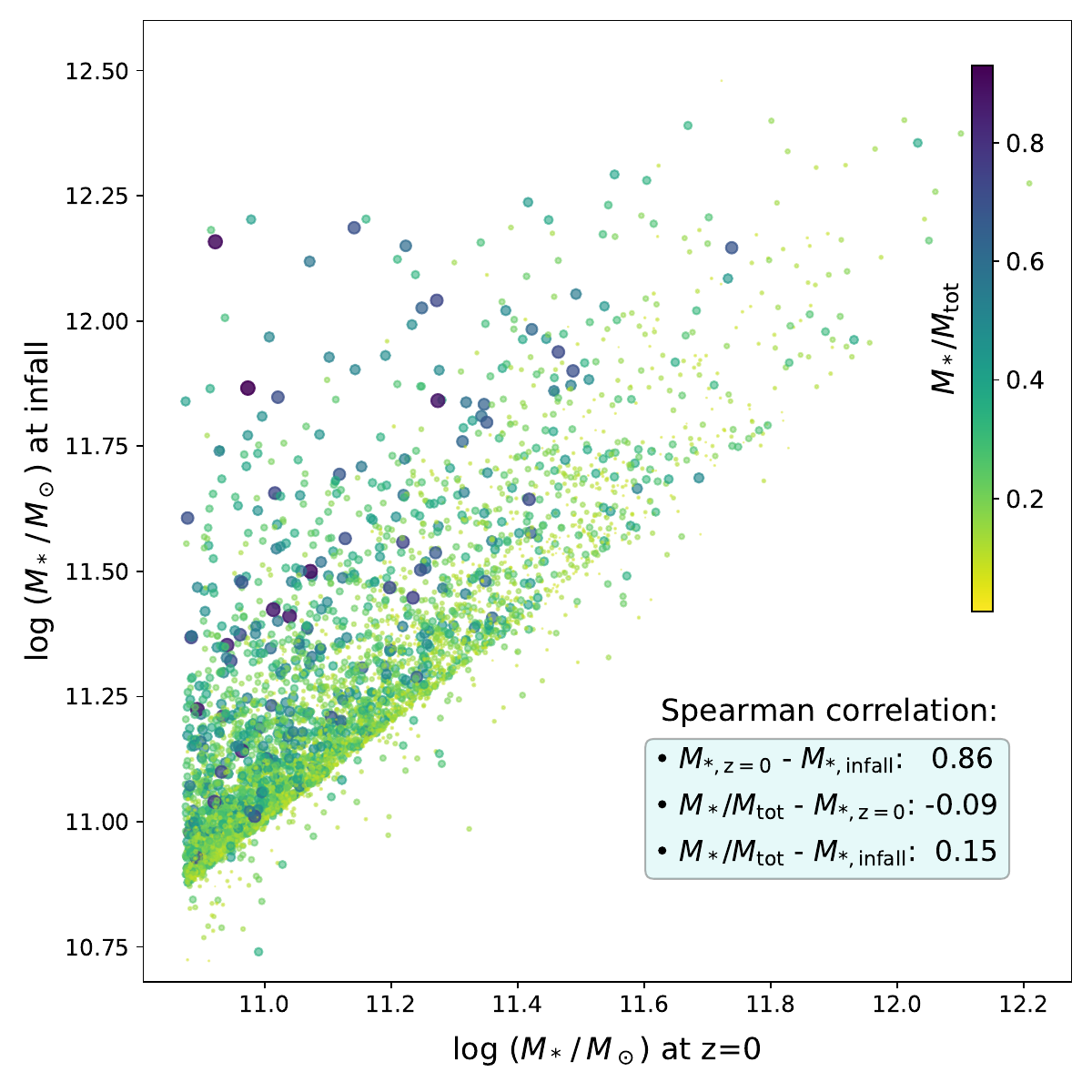}
  \caption{Stellar mass at infall as a function of the stellar mass today. The colours (and size) indicate the stellar-to-total mass ratio, as seen in the central colourbar, with darker (and larger) dots depicting galaxies with a higher value of $M_*/M_\mathrm{tot}$, i.e., with less DM.}
  \label{fig:mstar_infall}
\end{figure}

We have previously seen how tidal stripping affects satellite galaxies living within a host galaxy cluster, and how the infall times of these galaxies, together with the number of orbits and minimum pericentre distance of these orbits can lead to significantly stripped DM halos. While this explains how galaxies become dark matter-deficient, we still need to address their other key characteristic: stellar mass. Previously in this section, we have selected the most massive satellite galaxies in \textsc{The Three Hundred} clusters at $z=0$. Now, in order to connect to their initial conditions, we compare the mass of the galaxies today with their mass at infall time.

In \Fig{fig:mstar_infall} we show the scatter plot between the stellar mass at infall and the stellar mass at $z=0$. Similarly to Figs.~\ref{fig:distance-orbits} and \ref{fig:orbits-zinfall}, the dots are coloured (and sized) by $M_*/M_\mathrm{tot}$, so that darker (and larger) dots indicate more dark matter-deficient objects. \Fig{fig:mstar_infall} shows that objects tend to accumulate at the lower mass regions of the plot (simply reflecting the shape of the stellar mass function) and close to the identity line, indicating that, in general, the stellar component of satellite galaxies remains largely intact after falling into the cluster. Although there is significant scatter, the stellar mass at infall and the stellar mass today show a very tight correlation, as indicated by their Spearman correlation coefficient.

\begin{figure}
\centering
  \hspace*{-0.3cm}
  \includegraphics[width=9.5cm]{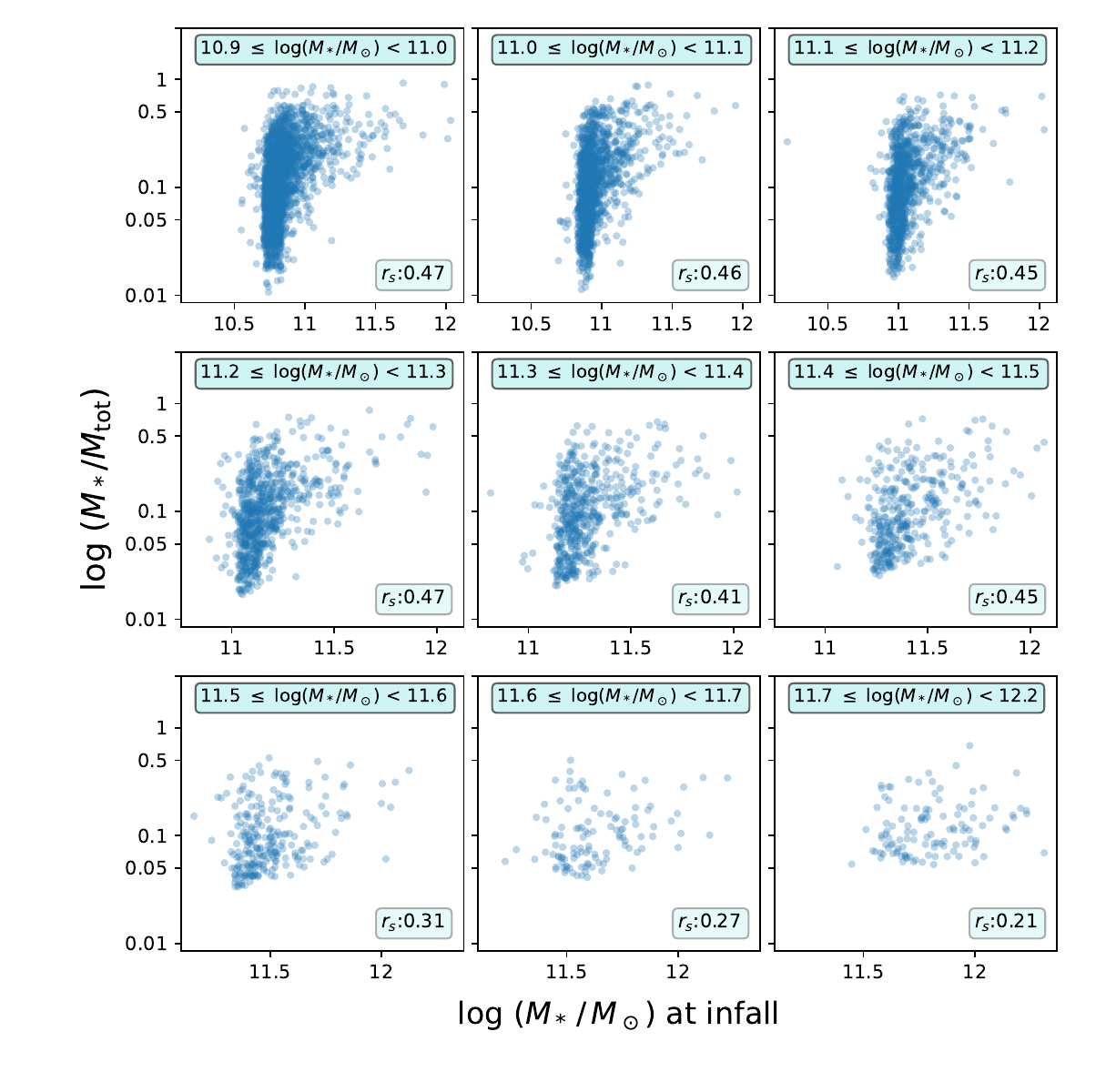}
  \vspace*{-0.8cm}
  \caption{Stellar-to-total mass ratio today as a function of the stellar mass at infall, separated in different bins for fixed values of the galaxy mass today. The values of the masses are indicated above. The Spearman correlation coefficient within each bin is shown in the bottom right corner.}
  \label{fig:mstar_infall_binning}
\end{figure}

Focusing now on the more dark matter-deficient galaxies in \Fig{fig:mstar_infall}, we can see that they are spread across different masses. However, as we saw before for the orbits, we now observe that, for a given mass today, darker dots seem to have a higher mass at infall than the lighter ones. This suggests that objects that are more massive (in stars) at infall tend to be more DM-stripped at $z=0$ than those less massive at their infall time. The Spearman correlation coefficients indicated in the plot further support this view. The values indicate that the stellar-to-total mass ratio today is more correlated with the mass at infall time than with the mass today. Although for the former ($M_*/M_\mathrm{tot}$ against $M_*$ at infall) the correlation is not very strong, we have checked that the correlation becomes stronger if considering bins with fixed mass today, reaching a Spearman correlation coefficient of $\sim 0.45$. This is illustrated in \Fig{fig:mstar_infall_binning}, where we show the relation between $M_*/M_\mathrm{tot}$ today and the stellar mass at infall, but for bins of fixed stellar mass today. The Spearman correlation coefficient is indicated for each bin, and it is clear that there is a significant correlation such that, for the same final mass, galaxies that were more massive at infall are more dark matter-deficient today. 

The mechanism described in the previous sections can also explain this scenario. If two galaxies end up with the same stellar mass at $z=0$, this means that the one that was more massive at infall has been stripped more since entering the cluster. But, since the dark matter is more easily stripped than the stars, more stellar stripping means even more dark matter stripping. So, after sufficient orbits and sufficiently close to the host centre, the galaxy can eventually become very dark matter-deficient as well as preserving a significant amount of its initial stellar content, that is, being still very massive.

To summarise, in this section we have shown that massive dark matter-deficient galaxies do exist in our simulations, but they require some specific conditions to be formed. They experienced a larger number of orbits and have on average smaller pericentre distances than average galaxies. At infall time, these objects are already very massive, with a concentrated stellar core, either being a single massive galaxy or the central galaxy of an infalling group. While the stars are more concentrated towards the centre, the dark matter halo is generally more extended, and so it is more easily stripped. When a galaxy undergoes many orbits, with pericentres very close to the host centre, this effect is intensified and can give rise to objects at $z=0$ as extreme as the examples presented here (see \Sec{sec:DMD-examples}).

\section{Conclusions} \label{sec:conclusions}

Motivated by the discovery of a massive, dark matter-deficient galaxy \citep[i.e. NGC 1277,][]{Comeron2023} in the Perseus galaxy cluster, we have used a set of 324 hydrodynamical re-simulations of galaxy clusters to search for galaxies that are similar in their stellar and dark matter contents. By analysing the stellar mass and stellar-to-total mass ratio of the satellite galaxies across the whole sample of 324 clusters, we find that there are some such galaxies with very low dark matter fractions and high stellar masses (see right panel in \Fig{fig:ratio-mstar}). The most extreme cases have stellar mass $M_* \sim 10^{11}M_{\odot}$ and stellar-to-total mass ratio $M_*/M_{\rm tot} \gtrsim 0.8$ at $z=0$, noting that they are all well resolved with $\geq$1000 (star) particles at redshift $z=0$, that is, at the end point of the stripping process. The existence of such objects is therefore not in contradiction with theoretical $\Lambda$CDM predictions. 

Although they are very rare, we have shown that massive dark matter-deficient galaxies can form naturally in a galaxy cluster environment. To avoid defining arbitrary thresholds that do not have any physical meaning, in this work we have not explicitly defined `dark matter-deficient' galaxies, but rather focused on the stellar-to-total mass ratio as our variable, studying its dependence on other galaxy properties. However, to have an approximate idea of the abundance of this kind of galaxies, we can make some selections. For instance, \citet{Montero-Dorta2022} define DM-deficient galaxies as those with stellar-to-DM ratio greater than 1. Selecting a similar threshold for DM-deficiency of $M_*/M_\mathrm{tot} \geq 0.5$ and a threshold for relatively massive galaxies of $M_* > 4\cdot 10^{10}M_{\odot}$, we find 302 massive dark matter deficient galaxies across our whole sample. For our whole galaxy cluster sample, that contains 324 clusters with an average of 200 satellite galaxies, this means around 0.5 per cent of the total satellite galaxy population. Setting further constraints to select the most extreme objects in our sample, we find only 9 galaxies with $M_*/M_\mathrm{tot} \geq 0.8$ and $M_* > 7.5 \cdot 10^{10}M_{\odot}$, each of them being in a different host, which further highlights the very rare nature of this kind of galaxies.



Regarding the formation of massive dark matter deficient galaxies, we showed here that the lack of dark matter in these galaxies is a result of the prolonged interaction with their host after infall. As they orbit, tidal interactions strip away their dark matter, leaving behind predominantly stellar material, which was more concentrated towards the centre and hence harder to strip. This is a generally accepted physical scenario for the mass loss of subhalos after infall into a particularly dense host environment and a natural outcome of hierarchical structure formation \citep[see e.g.][]{Mayer2002,Hayashi2003,Gill2004,Kravtsov2004,Penarrubia2008,Ogiya2018,Errani2020,Nusser2020,Green2021,Stuecker2023}. However, studying the whole population of massive satellite galaxies in \textsc{The Three Hundred} clusters, we find that more dark matter-deficient galaxies have a high number of orbits and with pericentre distances on average closer to the cluster centre.


Analysing the evolution of the galaxies in more detail, we further find that galaxies that are more dark matter-deficient also have on average higher infall redshifts than general galaxies. Moreover, when focusing on galaxies with similar infall times, that is, galaxies that have been inside the host for approximately the same time, we see that those that are more dark matter deficient have also conducted more orbits, which has enhanced their tidal stripping. 

Finally, in order to check for distinguishing features of the galaxies before infalling into the cluster, we compare the stellar masses of the galaxies at their own infall time. We show that the galaxies that end up being more dark matter-deficient today were amongst the most massive ones at their infall time, indicating that to obtain galaxies as extreme as the ones we obtain, an already very massive object at infall is required.



To sum up, we want to highlight that the purpose of this paper is to show that dark matter-deficient galaxies as massive as $M_* \sim 10^{11}$ M$_\odot$ can exist within a $\Lambda$CDM context, and thus are not in contradiction with theoretical predictions. We cannot claim that our proposed mechanism is responsible for the formation of a galaxy such as NGC 1277, but rather remark that, if certain conditions -- described above -- in their evolution are satisfied, galaxies can naturally evolve into massive dark matter-deficient galaxies without the need for going beyond the standard galaxy formation and evolution model. We also show that the final properties of these galaxies are a consequence of astrophysical processes, that is, tidal stripping, and not a numerical artefact, as it was the case in previous works that addressed the same issue. We further provide a study of the consistency of our results in Appendix~\ref{appendix:convergence}.

\begin{acknowledgements}
    This work has been made possible by \textsc{The Three Hundred} (\url{https://the300-project.org}) collaboration. ACS, AK, and WC are supported by the Ministerio de Ciencia e Innovaci\'{o}n (MICINN) under research grant PID2021-122603NB-C21. AK further thanks Us3 for cantaloop. FB acknowledges support from the project PID2020-116188GA-I00, funded by MICIU/AEI /10.13039/501100011033. WC is further supported by the STFC AGP Grant ST/V000594/1 and the Atracci\'{o}n de Talento Contract no. 2020-T1/TIC-19882 granted by the Comunidad de Madrid in Spain. He also thanks the science research grants from the China Manned Space Project and ERC: HORIZON-TMA-MSCA-SE for supporting the LACEGAL-III project with grant number 101086388. The simulations used in this paper have been performed in the MareNostrum Supercomputer at the Barcelona Supercomputing Center, thanks to CPU time granted by the Red Espa\~{n}ola de Supercomputaci\'on. As part of \textsc{The Three Hundred} project, this work has received financial support from the European Union’s Horizon 2020 Research and Innovation programme under the Marie Sklodowskaw-Curie grant agreement number 734374, the LACEGAL project.
\end{acknowledgements}


\bibliographystyle{aa}
\bibliography{archive}

\begin{thebibliography}{70}
\expandafter\ifx\csname natexlab\endcsname\relax\def\natexlab#1{#1}\fi

\bibitem[{{Allgood} {et~al.}(2006){Allgood}, {Flores}, {Primack}, {Kravtsov}, {Wechsler}, {Faltenbacher}, \& {Bullock}}]{Allgood2006}
{Allgood}, B., {Flores}, R.~A., {Primack}, J.~R., {et~al.} 2006, \mnras, 367, 1781

\bibitem[{{Barnes} \& {Hernquist}(1992)}]{Barnes-Hernquist1992}
{Barnes}, J.~E. \& {Hernquist}, L. 1992, \nat, 360, 715

\bibitem[{{Beck} {et~al.}(2016){Beck}, {Murante}, {Arth}, {Remus}, {Teklu}, {Donnert}, {Planelles}, {Beck}, {F{\"o}rster}, {Imgrund}, {Dolag}, \& {Borgani}}]{Beck2016}
{Beck}, A.~M., {Murante}, G., {Arth}, A., {et~al.} 2016, \mnras, 455, 2110

\bibitem[{{Biffi} {et~al.}(2017){Biffi}, {Planelles}, {Borgani}, {Fabjan}, {Rasia}, {Murante}, {Tornatore}, {Dolag}, {Granato}, {Gaspari}, \& {Beck}}]{Biffi2017}
{Biffi}, V., {Planelles}, S., {Borgani}, S., {et~al.} 2017, \mnras, 468, 531

\bibitem[{{Choi} {et~al.}(2009){Choi}, {Weinberg}, \& {Katz}}]{Choi2009}
{Choi}, J.-H., {Weinberg}, M.~D., \& {Katz}, N. 2009, \mnras, 400, 1247

\bibitem[{{Comer{\'o}n} {et~al.}(2023){Comer{\'o}n}, {Trujillo}, {Cappellari}, {Buitrago}, {Gardu{\~n}o}, {Zaragoza-Cardiel}, {Zinchenko}, {Lara-L{\'o}pez}, {Ferr{\'e}-Mateu}, \& {Dib}}]{Comeron2023}
{Comer{\'o}n}, S., {Trujillo}, I., {Cappellari}, M., {et~al.} 2023, \aap, 675, A143

\bibitem[{{Contreras-Santos} {et~al.}(2022){Contreras-Santos}, {Knebe}, {Pearce}, {Haggar}, {Gray}, {Cui}, {Yepes}, {De Petris}, {De Luca}, {Power}, {Mostoghiu}, {Nuza}, \& {Hoeft}}]{Contreras-Santos2022a}
{Contreras-Santos}, A., {Knebe}, A., {Pearce}, F., {et~al.} 2022, \mnras, 511, 2897

\bibitem[{{Cui} {et~al.}(2022){Cui}, {Dave}, {Knebe}, {Rasia}, {Gray}, {Pearce}, {Power}, {Yepes}, {Anbajagane}, {Ceverino}, {Contreras-Santos}, {de Andres}, {De Petris}, {Ettori}, {Haggar}, {Li}, {Wang}, {Yang}, {Borgani}, {Dolag}, {Zu}, {Kuchner}, {Ca{\~n}as}, {Ferragamo}, \& {Gianfagna}}]{Cui2022}
{Cui}, W., {Dave}, R., {Knebe}, A., {et~al.} 2022, \mnras, 514, 977

\bibitem[{{Cui} {et~al.}(2018){Cui}, {Knebe}, {Yepes}, {Pearce}, {Power}, {Dave}, {Arth}, {Borgani}, {Dolag}, {Elahi}, {Mostoghiu}, {Murante}, {Rasia}, {Stoppacher}, {Vega-Ferrero}, {Wang}, {Yang}, {Benson}, {Cora}, {Croton}, {Sinha}, {Stevens}, {Vega-Mart{\'\i}nez}, {Arthur}, {Baldi}, {Ca{\~n}as}, {Cialone}, {Cunnama}, {De Petris}, {Durando}, {Ettori}, {Gottl{\"o}ber}, {Nuza}, {Old}, {Pilipenko}, {Sorce}, \& {Welker}}]{Cui2018}
{Cui}, W., {Knebe}, A., {Yepes}, G., {et~al.} 2018, \mnras, 480, 2898

\bibitem[{{Danieli} {et~al.}(2020){Danieli}, {van Dokkum}, {Abraham}, {Conroy}, {Dolphin}, \& {Romanowsky}}]{Danieli2020}
{Danieli}, S., {van Dokkum}, P., {Abraham}, R., {et~al.} 2020, \apjl, 895, L4

\bibitem[{{Dolag} {et~al.}(2009){Dolag}, {Borgani}, {Murante}, \& {Springel}}]{Dolag2009}
{Dolag}, K., {Borgani}, S., {Murante}, G., \& {Springel}, V. 2009, \mnras, 399, 497

\bibitem[{{Drakos} {et~al.}(2020){Drakos}, {Taylor}, \& {Benson}}]{Drakos2020}
{Drakos}, N.~E., {Taylor}, J.~E., \& {Benson}, A.~J. 2020, \mnras, 494, 378

\bibitem[{{Duc} {et~al.}(2004){Duc}, {Bournaud}, \& {Masset}}]{Duc04}
{Duc}, P.~A., {Bournaud}, F., \& {Masset}, F. 2004, \aap, 427, 803

\bibitem[{{Errani} \& {Navarro}(2021)}]{Errani2021}
{Errani}, R. \& {Navarro}, J.~F. 2021, \mnras, 505, 18

\bibitem[{{Errani} \& {Pe{\~n}arrubia}(2020)}]{Errani2020}
{Errani}, R. \& {Pe{\~n}arrubia}, J. 2020, \mnras, 491, 4591

\bibitem[{{Gill} {et~al.}(2004){Gill}, {Knebe}, \& {Gibson}}]{Gill2004}
{Gill}, S. P.~D., {Knebe}, A., \& {Gibson}, B.~K. 2004, \mnras, 351, 399

\bibitem[{{Golini} {et~al.}(2024){Golini}, {Montes}, {Carrasco}, {Rom{\'a}n}, \& {Trujillo}}]{Golini2024}
{Golini}, G., {Montes}, M., {Carrasco}, E.~R., {Rom{\'a}n}, J., \& {Trujillo}, I. 2024, \aap, 684, A99

\bibitem[{{Green} {et~al.}(2021){Green}, {van den Bosch}, \& {Jiang}}]{Green2021}
{Green}, S.~B., {van den Bosch}, F.~C., \& {Jiang}, F. 2021, \mnras, 503, 4075

\bibitem[{{Hayashi} {et~al.}(2003){Hayashi}, {Navarro}, {Taylor}, {Stadel}, \& {Quinn}}]{Hayashi2003}
{Hayashi}, E., {Navarro}, J.~F., {Taylor}, J.~E., {Stadel}, J., \& {Quinn}, T. 2003, \apj, 584, 541

\bibitem[{{Jackson} {et~al.}(2021){Jackson}, {Kaviraj}, {Martin}, {Devriendt}, {Slyz}, {Silk}, {Dubois}, {Yi}, {Pichon}, {Volonteri}, {Choi}, {Kimm}, {Kraljic}, \& {Peirani}}]{Jackson2021}
{Jackson}, R.~A., {Kaviraj}, S., {Martin}, G., {et~al.} 2021, \mnras, 502, 1785

\bibitem[{{Jing} {et~al.}(2019){Jing}, {Wang}, {Li}, {Liao}, {Wang}, {Guo}, \& {Gao}}]{Jing2019}
{Jing}, Y., {Wang}, C., {Li}, R., {et~al.} 2019, \mnras, 488, 3298

\bibitem[{{Joshi} {et~al.}(2019){Joshi}, {Parker}, {Wadsley}, \& {Keller}}]{Joshi2019}
{Joshi}, G.~D., {Parker}, L.~C., {Wadsley}, J., \& {Keller}, B.~W. 2019, \mnras, 483, 235

\bibitem[{{Keim} {et~al.}(2022){Keim}, {van Dokkum}, {Danieli}, {Lokhorst}, {Li}, {Shen}, {Abraham}, {Chen}, {Gilhuly}, {Liu}, {Merritt}, {Miller}, {Pasha}, \& {Polzin}}]{Keim2022}
{Keim}, M.~A., {van Dokkum}, P., {Danieli}, S., {et~al.} 2022, \apj, 935, 160

\bibitem[{{Klypin} {et~al.}(2016){Klypin}, {Yepes}, {Gottl{\"o}ber}, {Prada}, \& {He{\ss}}}]{Klypin2016}
{Klypin}, A., {Yepes}, G., {Gottl{\"o}ber}, S., {Prada}, F., \& {He{\ss}}, S. 2016, \mnras, 457, 4340

\bibitem[{{Knebe} {et~al.}(2011){Knebe}, {Libeskind}, {Doumler}, {Yepes}, {Gottl{\"o}ber}, \& {Hoffman}}]{Knebe2011}
{Knebe}, A., {Libeskind}, N.~I., {Doumler}, T., {et~al.} 2011, \mnras, 417, L56

\bibitem[{{Knebe} {et~al.}(2013){Knebe}, {Libeskind}, {Pearce}, {Behroozi}, {Casado}, {Dolag}, {Dominguez-Tenreiro}, {Elahi}, {Lux}, {Muldrew}, \& {Onions}}]{Knebe2013}
{Knebe}, A., {Libeskind}, N.~I., {Pearce}, F., {et~al.} 2013, \mnras, 428, 2039

\bibitem[{{Knebe} {et~al.}(2006){Knebe}, {Power}, {Gill}, \& {Gibson}}]{Knebe2006}
{Knebe}, A., {Power}, C., {Gill}, S. P.~D., \& {Gibson}, B.~K. 2006, \mnras, 368, 741

\bibitem[{{Knollmann} \& {Knebe}(2009)}]{KnollmannKnebe2009}
{Knollmann}, S.~R. \& {Knebe}, A. 2009, \apjs, 182, 608

\bibitem[{{Kravtsov} {et~al.}(2004){Kravtsov}, {Gnedin}, \& {Klypin}}]{Kravtsov2004}
{Kravtsov}, A.~V., {Gnedin}, O.~Y., \& {Klypin}, A.~A. 2004, \apj, 609, 482

\bibitem[{{Lee} {et~al.}(2021){Lee}, {Shin}, \& {Kim}}]{Lee2021}
{Lee}, J., {Shin}, E.-j., \& {Kim}, J.-h. 2021, \apjl, 917, L15

\bibitem[{{Lewis} {et~al.}(2020){Lewis}, {Brewer}, \& {Wan}}]{Lewis20}
{Lewis}, G.~F., {Brewer}, B.~J., \& {Wan}, Z. 2020, \mnras, 491, L1

\bibitem[{{Libeskind} {et~al.}(2011){Libeskind}, {Knebe}, {Hoffman}, {Gottl{\"o}ber}, \& {Yepes}}]{Libeskind2011}
{Libeskind}, N.~I., {Knebe}, A., {Hoffman}, Y., {Gottl{\"o}ber}, S., \& {Yepes}, G. 2011, \mnras, 418, 336

\bibitem[{{Macci{\`o}} {et~al.}(2021){Macci{\`o}}, {Prats}, {Dixon}, {Buck}, {Waterval}, {Arora}, {Courteau}, \& {Kang}}]{Maccio2021}
{Macci{\`o}}, A.~V., {Prats}, D.~H., {Dixon}, K.~L., {et~al.} 2021, \mnras, 501, 693

\bibitem[{{Mayer} {et~al.}(2002){Mayer}, {Moore}, {Quinn}, {Governato}, \& {Stadel}}]{Mayer2002}
{Mayer}, L., {Moore}, B., {Quinn}, T., {Governato}, F., \& {Stadel}, J. 2002, \mnras, 336, 119

\bibitem[{{Monelli} \& {Trujillo}(2019)}]{Monelli19}
{Monelli}, M. \& {Trujillo}, I. 2019, \apjl, 880, L11

\bibitem[{{Montero-Dorta} {et~al.}(2022){Montero-Dorta}, {Rodriguez}, {Artale}, {Smith}, \& {Chaves-Montero}}]{Montero-Dorta2022}
{Montero-Dorta}, A.~D., {Rodriguez}, F., {Artale}, M.~C., {Smith}, R., \& {Chaves-Montero}, J. 2022, arXiv e-prints, arXiv:2212.12090

\bibitem[{{Montes} {et~al.}(2020){Montes}, {Infante-Sainz}, {Madrigal-Aguado}, {Rom{\'a}n}, {Monelli}, {Borlaff}, \& {Trujillo}}]{Montes20}
{Montes}, M., {Infante-Sainz}, R., {Madrigal-Aguado}, A., {et~al.} 2020, \apj, 904, 114

\bibitem[{{Montes} {et~al.}(2021){Montes}, {Trujillo}, {Infante-Sainz}, {Monelli}, \& {Borlaff}}]{Montes21}
{Montes}, M., {Trujillo}, I., {Infante-Sainz}, R., {Monelli}, M., \& {Borlaff}, A.~S. 2021, \apj, 919, 56

\bibitem[{{Murante} {et~al.}(2010){Murante}, {Monaco}, {Giovalli}, {Borgani}, \& {Diaferio}}]{Murante2010}
{Murante}, G., {Monaco}, P., {Giovalli}, M., {Borgani}, S., \& {Diaferio}, A. 2010, \mnras, 405, 1491

\bibitem[{{Newton} {et~al.}(2022){Newton}, {Libeskind}, {Knebe}, {S{\'a}nchez-Conde}, {Sorce}, {Pilipenko}, {Steinmetz}, {Pakmor}, {Tempel}, {Hoffman}, \& {Vogelsberger}}]{Newton2022}
{Newton}, O., {Libeskind}, N.~I., {Knebe}, A., {et~al.} 2022, \mnras, 514, 3612

\bibitem[{{Niemiec} {et~al.}(2019){Niemiec}, {Jullo}, {Giocoli}, {Limousin}, \& {Jauzac}}]{Niemiec2019}
{Niemiec}, A., {Jullo}, E., {Giocoli}, C., {Limousin}, M., \& {Jauzac}, M. 2019, \mnras, 487, 653

\bibitem[{{Nusser}(2020)}]{Nusser2020}
{Nusser}, A. 2020, \apj, 893, 66

\bibitem[{{Ogiya}(2018)}]{Ogiya2018}
{Ogiya}, G. 2018, \mnras, 480, L106

\bibitem[{{Ogiya} {et~al.}(2022){Ogiya}, {van den Bosch}, \& {Burkert}}]{Ogiya22}
{Ogiya}, G., {van den Bosch}, F.~C., \& {Burkert}, A. 2022, \mnras, 510, 2724

\bibitem[{{Osipova} {et~al.}(2023){Osipova}, {Pilipenko}, {Gottl{\"o}ber}, {Libeskind}, {Newton}, {Sorce}, \& {Yepes}}]{Osipova2023}
{Osipova}, A., {Pilipenko}, S., {Gottl{\"o}ber}, S., {et~al.} 2023, Physics of the Dark Universe, 42, 101328

\bibitem[{{Pe{\~n}arrubia} {et~al.}(2008){Pe{\~n}arrubia}, {Navarro}, \& {McConnachie}}]{Penarrubia2008}
{Pe{\~n}arrubia}, J., {Navarro}, J.~F., \& {McConnachie}, A.~W. 2008, \apj, 673, 226

\bibitem[{{Planelles} {et~al.}(2017){Planelles}, {Fabjan}, {Borgani}, {Murante}, {Rasia}, {Biffi}, {Truong}, {Ragone-Figueroa}, {Granato}, {Dolag}, {Pierpaoli}, {Beck}, {Steinborn}, \& {Gaspari}}]{Planelles2017}
{Planelles}, S., {Fabjan}, D., {Borgani}, S., {et~al.} 2017, \mnras, 467, 3827

\bibitem[{{Ploeckinger} {et~al.}(2018){Ploeckinger}, {Sharma}, {Schaye}, {Crain}, {Schaller}, \& {Barber}}]{Ploeckinger2018}
{Ploeckinger}, S., {Sharma}, K., {Schaye}, J., {et~al.} 2018, \mnras, 474, 580

\bibitem[{{Rasia} {et~al.}(2015){Rasia}, {Borgani}, {Murante}, {Planelles}, {Beck}, {Biffi}, {Ragone-Figueroa}, {Granato}, {Steinborn}, \& {Dolag}}]{Rasia2015}
{Rasia}, E., {Borgani}, S., {Murante}, G., {et~al.} 2015, \apjl, 813, L17

\bibitem[{{Richings} {et~al.}(2020){Richings}, {Frenk}, {Jenkins}, {Robertson}, {Fattahi}, {Grand}, {Navarro}, {Pakmor}, {Gomez}, {Marinacci}, \& {Oman}}]{Richards2020}
{Richings}, J., {Frenk}, C., {Jenkins}, A., {et~al.} 2020, \mnras, 492, 5780

\bibitem[{{Santucci} {et~al.}(2022){Santucci}, {Brough}, {van de Sande}, {McDermid}, {van de Ven}, {Zhu}, {D'Eugenio}, {Bland-Hawthorn}, {Barsanti}, {Bryant}, {Croom}, {Davies}, {Green}, {Lawrence}, {Lorente}, {Owers}, {Poci}, {Richards}, {Thater}, \& {Yi}}]{Santucci22}
{Santucci}, G., {Brough}, S., {van de Sande}, J., {et~al.} 2022, \apj, 930, 153

\bibitem[{{Saulder} {et~al.}(2020){Saulder}, {Snaith}, {Park}, \& {Laigle}}]{Saulder2020}
{Saulder}, C., {Snaith}, O., {Park}, C., \& {Laigle}, C. 2020, \mnras, 491, 1278

\bibitem[{{Shen} {et~al.}(2023){Shen}, {van Dokkum}, \& {Danieli}}]{Shen2023}
{Shen}, Z., {van Dokkum}, P., \& {Danieli}, S. 2023, \apj, 957, 6

\bibitem[{{Shin} {et~al.}(2020){Shin}, {Jung}, {Kwon}, {Kim}, {Lee}, {Jo}, \& {Oh}}]{Shin2020}
{Shin}, E.-j., {Jung}, M., {Kwon}, G., {et~al.} 2020, \apj, 899, 25

\bibitem[{{Silk}(2019)}]{Silk2019}
{Silk}, J. 2019, \mnras, 488, L24

\bibitem[{{Smith} {et~al.}(2022){Smith}, {Calder{\'o}n-Castillo}, {Shin}, {Raouf}, \& {Ko}}]{Smith2022}
{Smith}, R., {Calder{\'o}n-Castillo}, P., {Shin}, J., {Raouf}, M., \& {Ko}, J. 2022, \aj, 164, 95

\bibitem[{{Smith} {et~al.}(2016){Smith}, {Choi}, {Lee}, {Rhee}, {Sanchez-Janssen}, \& {Yi}}]{Smith2016}
{Smith}, R., {Choi}, H., {Lee}, J., {et~al.} 2016, \apj, 833, 109

\bibitem[{{Springel}(2005)}]{Springel2005}
{Springel}, V. 2005, \mnras, 364, 1105

\bibitem[{{Springel} \& {Hernquist}(2003)}]{Springel2003}
{Springel}, V. \& {Hernquist}, L. 2003, \mnras, 339, 289

\bibitem[{{Srisawat} {et~al.}(2013){Srisawat}, {Knebe}, {Pearce}, {Schneider}, {Thomas}, {Behroozi}, {Dolag}, {Elahi}, {Han}, {Helly}, {Jing}, {Jung}, {Lee}, {Mao}, {Onions}, {Rodriguez-Gomez}, {Tweed}, \& {Yi}}]{Srisawat2013}
{Srisawat}, C., {Knebe}, A., {Pearce}, F.~R., {et~al.} 2013, \mnras, 436, 150

\bibitem[{{Steinborn} {et~al.}(2015){Steinborn}, {Dolag}, {Hirschmann}, {Prieto}, \& {Remus}}]{Steinborn2015}
{Steinborn}, L.~K., {Dolag}, K., {Hirschmann}, M., {Prieto}, M.~A., \& {Remus}, R.-S. 2015, \mnras, 448, 1504

\bibitem[{{St{\"u}cker} {et~al.}(2023){St{\"u}cker}, {Ogiya}, {Angulo}, {Aguirre-Santaella}, \& {S{\'a}nchez-Conde}}]{Stuecker2023}
{St{\"u}cker}, J., {Ogiya}, G., {Angulo}, R.~E., {Aguirre-Santaella}, A., \& {S{\'a}nchez-Conde}, M.~A. 2023, \mnras, 521, 4432

\bibitem[{{Tornatore} {et~al.}(2007){Tornatore}, {Borgani}, {Dolag}, \& {Matteucci}}]{Tornatore2007}
{Tornatore}, L., {Borgani}, S., {Dolag}, K., \& {Matteucci}, F. 2007, \mnras, 382, 1050

\bibitem[{{Trujillo} {et~al.}(2019){Trujillo}, {Beasley}, {Borlaff}, {Carrasco}, {Di Cintio}, {Filho}, {Monelli}, {Montes}, {Rom{\'a}n}, {Ruiz-Lara}, {S{\'a}nchez Almeida}, {Valls-Gabaud}, \& {Vazdekis}}]{Trujillo19}
{Trujillo}, I., {Beasley}, M.~A., {Borlaff}, A., {et~al.} 2019, \mnras, 486, 1192

\bibitem[{{van den Bosch} {et~al.}(2018){van den Bosch}, {Ogiya}, {Hahn}, \& {Burkert}}]{vandenBosch2018a}
{van den Bosch}, F.~C., {Ogiya}, G., {Hahn}, O., \& {Burkert}, A. 2018, \mnras, 474, 3043

\bibitem[{{van Dokkum} {et~al.}(2019){van Dokkum}, {Danieli}, {Abraham}, {Conroy}, \& {Romanowsky}}]{VanDokkum19}
{van Dokkum}, P., {Danieli}, S., {Abraham}, R., {Conroy}, C., \& {Romanowsky}, A.~J. 2019, \apjl, 874, L5

\bibitem[{{van Dokkum} {et~al.}(2018){van Dokkum}, {Danieli}, {Cohen}, {Merritt}, {Romanowsky}, {Abraham}, {Brodie}, {Conroy}, {Lokhorst}, {Mowla}, {O'Sullivan}, \& {Zhang}}]{VanDokkum18}
{van Dokkum}, P., {Danieli}, S., {Cohen}, Y., {et~al.} 2018, \nat, 555, 629

\bibitem[{{Wang} {et~al.}(2016){Wang}, {Pearce}, {Knebe}, {Schneider}, {Srisawat}, {Tweed}, {Jung}, {Han}, {Helly}, {Onions}, {Elahi}, {Thomas}, {Behroozi}, {Yi}, {Rodriguez-Gomez}, {Mao}, {Jing}, \& {Lin}}]{Wang2016}
{Wang}, Y., {Pearce}, F.~R., {Knebe}, A., {et~al.} 2016, \mnras, 459, 1554

\bibitem[{{White} \& {Rees}(1978)}]{White1978}
{White}, S.~D.~M. \& {Rees}, M.~J. 1978, \mnras, 183, 341

\bibitem[{{Yu} {et~al.}(2018){Yu}, {Ratra}, \& {Wang}}]{Yu2018}
{Yu}, H., {Ratra}, B., \& {Wang}, F.-Y. 2018, arXiv e-prints, arXiv:1809.05938

\end{thebibliography}

\appendix

\section{Evolution of host radius} \label{appendix:host-radius}

\begin{figure}[h]
\centering
  \includegraphics[width=9cm]{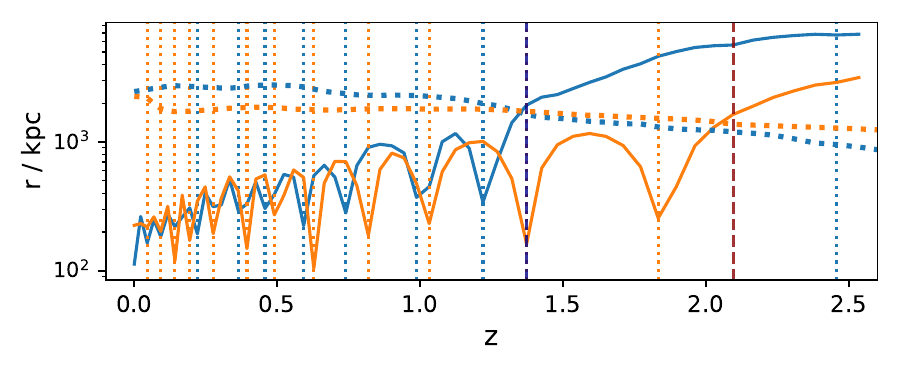}
  \caption{Bottom panel of \Fig{fig:evolution-rdist} in the main text, but with the distance in kiloparsecs, instead of normalising by the host radius. The horizontal dotted lines indicate the radius of each host.}
  \label{fig:appendix-radius}
\end{figure}

In Section \ref{sec:DMD-examples} we showed the evolution with time of two objects selected as the most extreme cases of massive dark matter-deficient galaxies. In the bottom panel of \Fig{fig:evolution-rdist} we displayed the evolution in the distance to the host centre, normalised by the radius of the host $R_{200}$. Here, in \Fig{fig:appendix-radius}, we show the same plot but without normalising by the radius, that is, showing the distance in kiloparsecs. As in \Fig{fig:evolution-rdist}, the two solid lines show the evolution for the two selected galaxies, with the vertical dotted lines indicating the pericentres and the dashed lines their respective infall times (see the main text in \Sec{sec:DMD-examples} for more details). The horizontal dotted lines indicate now the radius $R_{200}$ of the host, in kpc. It can be seen that, while the radius is clearly increasing over time, the shrinking of the orbits remains very clear. This confirms that the effect seen in \Fig{fig:evolution-rdist} is not a consequence of the radius of the host increasing, but rather of the galaxies becoming increasingly closer to the cluster centre.

\section{Convergence study}  \label{appendix:convergence}

Given that previous works that tried to find massive dark matter-deficient galaxies in simulations only found their appearance to be a numerical artefact, we also like to comment on a possible convergence study to show the consistency of our results. For instance, mass convergence means that, keeping the masses of the objects fixed but increasing the number of resolution elements used to resolve the object, the physical results will not change. Such experiments are extremely important and useful for as long as  additional physics do not depend on the number of resolution elements. However, for hydrodynamical simulations that include sub-grid modelling this is unfortunately not the case. For simulations such as the ones presented here, increasing the mass resolution will require the sub-grid physics parameters to be re-tuned to ensure the same kind of star formation and feedback. This prevents us from conducting a `conventional' convergence study.

\begin{figure}
\centering
  \includegraphics[width=8cm]{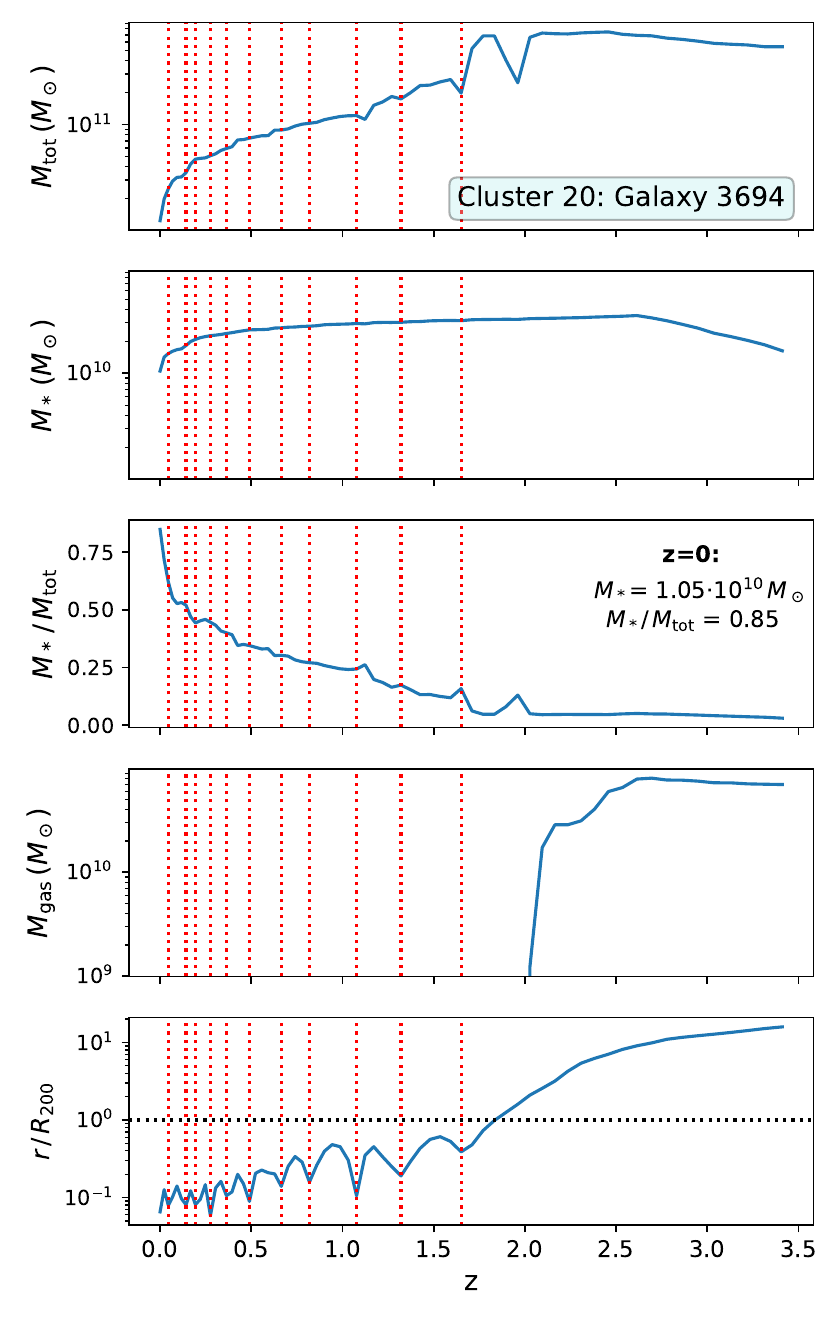}
  \caption{Replica of \Fig{fig:evolution-rdist} in the main text, but this time for a galaxy with $M_*\sim 10^{10}$ M$_\odot$ at $z=0$ (i.e. $\sim$ 10 times lower in stellar mass than the galaxies selected in the paper).}
  \label{fig:convergence1}
\end{figure}

Despite that, we nevertheless want to address this issue, approaching it from a different angle. In order to mimic a convergence study, we investigate the tidal stripping of galaxies less massive than the ones reported in the main text\footnote{We remind the reader that objects in Section \ref{sec:dmd-statistical} have $M_* \geq 7.5 \cdot 10^{10} M_\odot$.}. To have a situation as similar as possible to the dark matter-deficient objects discussed in the text, we focus on objects that are also dark matter-deficient at $z=0$. The idea is that these objects have fewer particles (but also lower masses) than the massive galaxies selected in the main text, and the question is if we can observe similar tidal stripping for them under the same orbital conditions. 
We confirm that galaxies selected according to this approach (i.e. galaxies with $M_*/M_{\rm tot}$ larger than the average and low stellar mass at $z=0$) in general had more orbits and closer pericentres than average galaxies. An example of this is shown in \Fig{fig:convergence1}, where we replicate \Fig{fig:evolution-rdist} but for a galaxy with $M_* \sim 10^{10}$ M$_\odot$. We find that it basically undergoes the same stripping as the better resolved counterparts.

As these is not a real convergence test in the classical sense, it should therefore also be interpreted with great care. It features objects from a different mass bin, and hence we should only take it as a `hint at convergence'. But we nevertheless want to stress again that all the galaxies selected in this work are well-resolved, with more than 1000 stellar particles at $z=0$ and far more at their infall time.

\section{Proof of concept}  \label{appendix:concept}
In the spirit of previous, controlled numerical experiments \citep[e.g.][amongst many others]{Hayashi2003, Penarrubia2008, Errani2020}, we also performed simulations of subhalos falling into and subsequently orbiting a host halo. This appendix should just be viewed as a supplementary toy model where we like to show that massive dark matter-deficient galaxies can be created via simple tidal stripping.

The host halo was modelled as an analytical Hernquist profile with a mass of $M_{200c}^{\rm host} = 10^{15}M_{\odot}$ and $R_{200c}=1626$~kpc. The scale radius was set to 281~kpc. The subhalo consisted of two components, a dark matter halo and a stellar bulge, both following a Hernquist profile with scale radii 27~kpc and 2.7~kpc, respectively. Each component was sampled with equal mass particles, that is, 900000 dark matter and 100000 star particles, amounting to a total of $M_{200c}^{\rm sub}=10^{12}M_{\odot}$. 80 per cent of the particles were inside $R_{200c}^{\rm sub}=163$~kpc. All initial conditions have been constructed using GalIC and the simulations have been run using Gadget-4 for 30 billion years resulting in 512 snapshots per configuration. The subhalo has been positioned on four distinct stable orbits with varying pericentre $R_{p} = [0.01, 0.1, 0.2, 0.4] R_{200c}^{\rm host}$, infalling from $R_{200c}^{\rm host}$.

\begin{figure}
\centering
  \includegraphics[width=8.5cm]{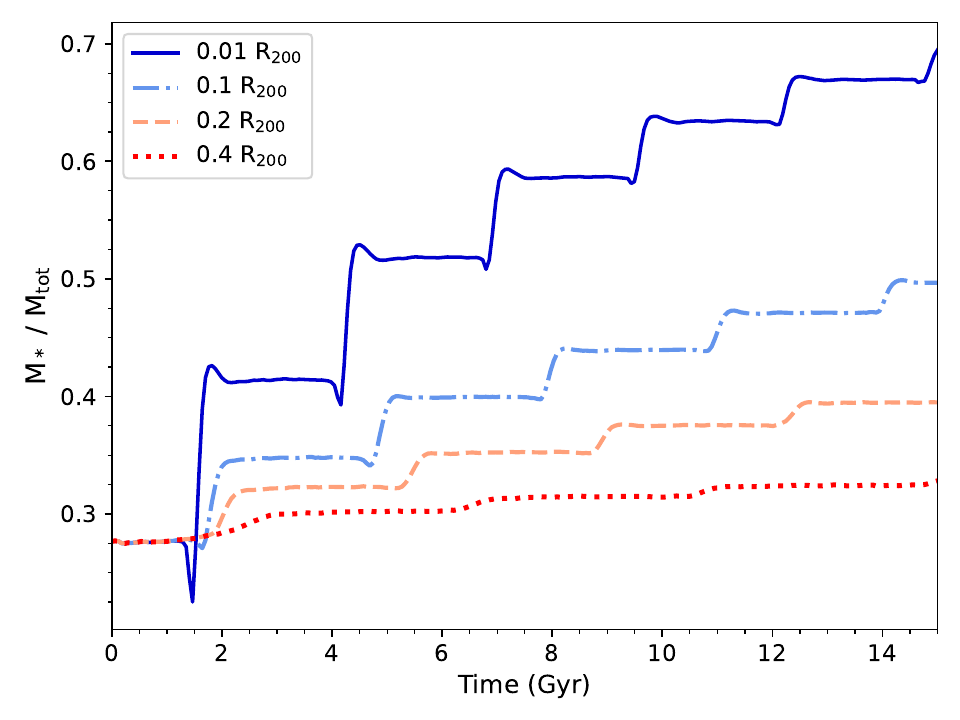}
  \caption{Temporal evolution of the stellar-to-halo mass ratio -- as measured inside the Hernquist scale-radius -- of a numerically modelled subhalo orbiting inside an analytical host halo.}
  \label{fig:concept}
\end{figure}

In \Fig{fig:concept} we show the temporal evolution up to 15 Gyrs of the stellar-to-halo mass ratio (as measured within the scale-radius of the subhalo) of the orbiting subhalo. We can clearly see how the ratio substantially increases for those orbits that bring the subhalo close to the host centre: the dark matter gets sequentially stripped while the stellar component is hardly affected by the tides. This can be substantiated by looking at the evolution of the actual mass components, but we decided to omit such a plot here as the ratio more clearly shows the transformation into a massive dark matter-deficient galaxy.

We acknowledge that this experiment is by no means trying to re-model one of the objects presented in the main text. It rather serves as a `proof of concept' that it is possible to construct initial configurations of dark matter and stars for objects in the same mass range as discussed in the paper, leading to a substantial increase in stellar-to-halo mass ratio after several orbits and when closely approaching the host's centre. 

\label{lastpage}
\end{document}